\documentclass[a4paper, 11pt]{article}
\pdfoutput=1

\usepackage{latexsym, amsmath, amsfonts, amssymb}
\usepackage{tikz}
\usetikzlibrary{decorations.pathmorphing, decorations.markings}
\usepackage{mathrsfs}
\usepackage[american]{babel}
\usepackage{graphicx}
\usepackage{bbm}
\usepackage{cite}

\usepackage[colorlinks=true, citecolor=blue!90!black, linkcolor=blue!90!black, linktocpage=true, urlcolor=red!70!black]{hyperref}

\renewcommand{\baselinestretch}{1.2}
\newcommand{\Vol}{\mathrm{Vol}}

\newcommand{\ds}{\displaystyle}
\newcommand{\wt}{\widetilde}

\newcommand{\matht}[1]{{\ensuremath{\boldsymbol{#1}}}}
\newcommand{\tps}[2]{\texorpdfstring{#1}{#2}}

\newcommand{\eg}{\textit{e.g.}}

\newcommand{\ie}{\textit{i.e.}}

\numberwithin{equation}{section}

\newcommand{\nn}{\nonumber}

\newcommand{\be}{\begin{equation}} \newcommand{\ee}{\end{equation}}
\newcommand{\bea}{\begin{equation} \begin{aligned}} \newcommand{\eea}{\end{aligned} \end{equation}}

\newcommand{\cA}{\mathcal{A}}
\newcommand{\cB}{\mathcal{B}}
\newcommand{\cC}{\mathcal{C}}

\newcommand{\cF}{\mathcal{F}}
\newcommand{\cG}{\mathcal{G}}
\newcommand{\cH}{\mathcal{H}}

\newcommand{\cL}{\mathcal{L}}

\newcommand{\cM}{\mathcal{M}}

\newcommand{\cO}{\mathcal{O}}
\newcommand{\cP}{\mathcal{P}}

\newcommand{\cT}{\mathcal{T}}

\newcommand{\cZ}{\mathcal{Z}}

\newcommand{\bC}{\mathbb{C}}

\newcommand{\bP}{\mathbb{P}}
\newcommand{\bQ}{\mathbb{Q}}
\newcommand{\bR}{\mathbb{R}}

\newcommand{\bZ}{\mathbb{Z}}

\newcommand{\fg}{\mathfrak{g}}

\newcommand{\fR}{\mathfrak{R}}

\newcommand{\sT}{\mathsf{T}}

\newcommand{\unit}{\mathbbm{1}}

\def\repa{\raise4pt\hbox{$\square$}\mkern-14mu\raise-4pt\hbox{$\square$}}
\def\repab{\overline{\raise4pt\hbox{$\square$}\mkern-14mu\raise-4pt\hbox{$\square$}\mkern-1mu}}

\DeclareMathOperator{\Tr}{Tr}

\begin{document}

\thispagestyle{empty}
\fontsize{12pt}{20pt}
\begin{flushright}
	SISSA 15/2024/FISI
\end{flushright}
\vspace{13mm}
\begin{center}
	{\huge Holographic duals of symmetry broken phases}
	\\[13mm]
    	{\large Andrea Antinucci$^{a,\, b}$, \, Francesco Benini$^{a,\, b,\, c}$ and Giovanni Rizi$^{a,\, b}$}
	
	\bigskip
	{\it
		$^a$ SISSA, Via Bonomea 265, 34136 Trieste, Italy \\[.0em]
		$^b$ INFN, Sezione di Trieste, Via Valerio 2, 34127 Trieste, Italy \\[.6em]
		$^c$ ICTP, Strada Costiera 11, 34151 Trieste, Italy}
\end{center}

\bigskip

\begin{abstract}
\noindent
We explore a novel interpretation of Symmetry Topological Field Theories (SymTFTs) as theories of gravity, proposing a holographic duality where the bulk SymTFT (with the gauging of a suitable Lagrangian algebra) is dual to the universal effective field theory (EFT) that describes spontaneous symmetry breaking on the boundary. We test this conjecture in various dimensions and with many examples involving different continuous symmetry structures, including non-Abelian and non-invertible symmetries, as well as higher groups. For instance, we find that many Abelian SymTFTs are dual to free theories of Goldstone bosons or generalized Maxwell fields, while non-Abelian SymTFTs relate to non-linear sigma models with target spaces defined by the symmetry groups. We also extend our analysis to include the non-invertible $\bQ/\bZ$ axial symmetry, finding it to be dual to axion-Maxwell theory, and a non-Abelian 2-group structure in four dimensions, deriving a new parity-violating interaction that has implications for the low-energy dynamics of $U(N)$ QCD.
\end{abstract}

\newpage
\pagenumbering{arabic}
\setcounter{page}{1}
\setcounter{footnote}{0}
\renewcommand{\thefootnote}{\arabic{footnote}}

{\renewcommand{\baselinestretch}{.88} \parskip=0pt
\setcounter{tocdepth}{2}
\tableofcontents}

%%%%%%%%%%%%%%%%%%%%%%%%%%%%%%%%%%%%%%%%%%%%%%%%%%%%

\section{Introduction}
\label{sec: intro}

A profound insight by E.~Witten is that Topological Quantum Field Theories (TQFTs), due to their general covariance, can be seen as theories of quantum gravity \cite{Witten:1988ze}. Unlike in more conventional examples, general covariance is not achieved by integrating over metrics but rather by not introducing them at all. Consequently, these theories lack any semiclassical description involving weakly interacting gravitons. In traditional gravitational theories, one selects a background metric and expands around it, thereby breaking general covariance spontaneously. Therefore, TQFTs can be viewed as theories of quantum gravity with unbroken general covariance --- where gravitons are, in a certain sense, confined.

This old story requires some important refinements. A full quantum-gravity theory should not depend on the background topology. TQFTs, on the other hand, are sensitive to spacetime topology through their global symmetries, broadly defined in terms of their topological operators \cite{Gaiotto:2014kfa}, which are expected to form some higher category \cite{Kapustin:2010ta, Bhardwaj:2017xup, Chang:2018iay, Choi:2021kmx, Kaidi:2021xfk, Roumpedakis:2022aik, Bhardwaj:2022yxj, Antinucci:2022eat, Freed:2022qnc, Argurio:2024oym}. One way to achieve such an independence is to sum over all topologies, which can be done in low dimensions \cite{Maloney:2007ud, Saad:2018bqo, Stanford:2019vob, Maloney:2020nni, Afkhami-Jeddi:2020ezh}. Alternatively, one can use TQFTs that do not even depend on topology \cite{Benini:2022hzx}, hence that are free of global symmetries and then trivial (or invertible) \cite{Freed:2014eja, Freed:2016rqq}. These can be obtained by gauging a maximal non-anomalous set of topological defects, called a \emph{Lagrangian algebra}, in a nontrivial TQFT. Not all TQFTs have Lagrangian algebras (the typical example is 3d Chern--Simons theory), but those that have them admit topological (or gapped) boundary conditions. In fact, given a Lagrangian algebra $\cL$, one can construct such a boundary condition as an interface between the TQFT and the gauged TQFT \cite{Davydov:2010kfz, Fuchs:2012dt, Freed:2020qfy, Kaidi:2021gbs}. Equivalently, the boundary condition is defined by allowing the defects inside $\cL$ to end on the boundary.

TQFTs with topological boundary conditions have recently gained attention for their role as Symmetry Topological Field Theories (SymTFTs) in the context of generalized symmetries (see \cite{Schafer-Nameki:2023jdn, Shao:2023gho, Brennan:2023mmt, Bhardwaj:2023kri} for reviews). SymTFTs are $(d+1)$-dimensional TQFTs $\cZ(\cC)$ associated with symmetry structures $\cC$ in $d$ dimensions, capturing all properties of the symmetries regardless of the specific QFT$_d$ realizing them \cite{Gaiotto:2014kfa, Gaiotto:2020iye, Apruzzi:2021nmk, Freed:2022qnc}. The TQFT $\cZ(\cC)$ is placed on a \emph{slab} with two boundaries. The left one supports the physical QFT$_d$ of interest, coupled with the bulk. The right one is the topological boundary condition that one is free to choose, determined by a Lagrangian algebra $\cL$. Defects inside $\cL$ become trivial on the topological boundary, while all other ones (modulo those inside $\cL$) give rise to topological operators of the symmetry $\cC$, after the slab is squeezed. The endpoints of defects inside $\cL$ inherit a braiding with the generators of $\cC$ from the bulk braiding, hence they become the charges of the symmetry \cite{Bhardwaj:2023wzd, Bhardwaj:2023ayw}.  SymTFT has been shown to be a powerful tool for studying global symmetries, also of non-invertible type \cite{Apruzzi:2022rei, Kaidi:2022cpf, Antinucci:2022vyk, Bashmakov:2022uek, Antinucci:2022cdi} and their anomalies \cite{Kaidi:2023maf, Zhang:2023wlu, Antinucci:2023ezl, Cordova:2023bja, Putrov:2024uor}, as well as to characterize phases \cite{ Wen:2022tkg, Wen:2023otf, Bhardwaj:2023idu, Bhardwaj:2023fca, Antinucci:2024ltv}.

Although originally restricted to finite symmetries, the framework has been recently extended to continuous symmetries \cite{Antinucci:2024zjp, Brennan:2024fgj, Bonetti:2024cjk}.%
\footnote{See \cite{Apruzzi:2024htg} for a different proposal involving non-topological theories.}
The prize to pay is to introduce a new type of TQFTs with gauge fields valued in both $U(1)$ and $\bR$, and to have a continuous and/or non-compact spectrum of operators, thus going beyond the standard TQFTs well studied by mathematicians (we provide a more precise mathematical definition in Appendix~\ref{app:TQFT}). This idea has been shown to be applicable to all possible non-finite and continuous symmetries, with or without anomalies, possibly with higher-group structures, and even including non-invertible and non-Abelian symmetries. By now the picture is that to \emph{any} possible symmetry structure $\cC$ in $d$ dimensions one can canonically associate a $(d+1)$-dimensional TQFT $\cZ(\cC)$.

Our aim here is to give a different interpretation to these TQFTs $\cZ(\cC)$, not as SymTFTs but as theories of gravity. More precisely, we want to establish holographic dualities in which the bulk theory is a SymTFT. The main proposal of this paper is the following:
{\setlength{\leftmargini}{1.5em}
\begin{itemize}
\item Thought of as a theory of gravity, the SymTFT $\cZ(\cC)$ for a symmetry $\cC$ is the holographic dual to the universal effective field theory (EFT) that describes the spontaneous breaking of $\cC$.
\end{itemize}}\noindent
It is a general principle of quantum field theory that any theory with a certain continuous global symmetry that is spontaneously broken, in the far infrared (IR) flows to the same universal theory of Goldstone bosons \cite{Coleman:1969sm, Callan:1969sn}. This is roughly speaking always a sigma model, although the target space can be infinite dimensional (\eg, it is the classifying space $B^pG$ in the case of higher-form symmetries).%
\footnote{It is not clear to us how to make this precise for non-invertible symmetries, for instance for the $\bQ/\bZ$ chiral symmetry discovered in \cite{Choi:2022jqy, Cordova:2022ieu}.}
As for the SymTFT, this EFT is also canonically determined by the symmetry $\cC$ without any further information. For this reason, it is natural to expect that, even though they appear to be completely different objects --- a $(d+1)$-dimensional TQFT and a $d$-dimensional EFT --- the two can be somehow related as they both have the same input datum. We will prove by means of many examples that this correspondence is holography.

A crucial part of the story is the proper choice of boundary conditions. These will be non-topological and of the Dirichlet type for some combination of the bulk fields. Since bulk fields are gauge fields $A$, these boundary conditions break some gauge invariance, making it a global symmetry of the boundary theory. This agrees with the general principle in holography that boundary global symmetries correspond to bulk gauge fields. The non-triviality of the system really comes from the boundary conditions that, being non-topological, generate dynamics on the boundary. The boundary theory can be thought of as a theory of edge modes. Our setup has several similarities with, and may be understood as a generalization of, the Chern--Simons/WZW correspondence \cite{Witten:1988hf, Elitzur:1989nr} and its reinterpretation as a full-flagged holographic duality by means of bulk anyon condensation \cite{Benini:2022hzx}.%
\footnote{See also \cite{Gukov:2004id} for earlier work, as well as \cite{Aharony:2023zit, Barbar:2023ncl} for a related approach.}

We find that for the simple Abelian TQFTs introduced in \cite{Brennan:2024fgj, Antinucci:2024zjp} as the SymTFTs for $U(1)$, the dual boundary theory is the free theory of an $S^1$ Goldstone boson, or generalized Maxwell field when the symmetry is of higher form. More precisely, these boundary theories have topological sectors (\eg, winding for a compact scalar, or magnetic fluxes for a photon), and the nontrivial TQFT without gauging the Lagrangian algebra is only dual to a fixed topological sector. The latter is not a physical theory and is the non-chiral analog of the conformal blocks in the CS/WZW correspondence. The physical theory is obtained by summing over various topological sectors, and we will show that this sum is reproduced by the gauging of the Lagrangian algebra. These Abelian TQFTs have various interesting modifications describing chiral anomalies, higher groups, and non-invertible $\bQ/\bZ$ symmetries \cite{Antinucci:2024zjp}. We include all of them in our analysis, showing that their holographic duals are the theories describing the spontaneous breaking of the corresponding symmetries. In particular the SymTFT for the non-invertible chiral symmetry is the gravity dual to axion-Maxwell theory.

For non-Abelian continuous symmetries $G$, the SymTFT was also conjectured in \cite{Brennan:2024fgj, Antinucci:2024zjp} and further analyzed in\cite{Bonetti:2024cjk}. In the simplest case, it is a TQFT introduced many years ago by Horowitz \cite{Horowitz:1989ng} and is written in terms of a $G$ connection and a Lie-algebra-valued higher-form field in the adjoint of $G$. When employing this theory in our story, it proves to be the dual to a non-linear sigma model with target space $G$ at the boundary. For $d=4$ this in the pion Lagrangian describing the low-energy dynamics of massless QCD in the chiral symmetry breaking phase. We also show that including a term that describes an 't~Hooft anomaly we obtain a WZW term in the sigma model \cite{Witten:1983tw}. 

A particularly interesting example is that of a non-Abelian 2-group in 4d, mixing a non-Abelian continuous symmetry $G$ and a $U(1)$ 1-form symmetry \cite{Cordova:2018cvg}. The Goldstone theory for this symmetry structure was not determined before, and we use our holographic conjecture to derive it. It consists of a non-linear sigma model and a photon, coupled through a parity-violating interaction whose leading term is proportional to $k f_{abc} \epsilon^{\mu \nu \rho \sigma } A_\mu \, \partial_\nu \pi_a \, \partial_\rho \pi_b \, \partial_\sigma \pi_c $, where $\pi _a$ are the pions, $f_{abc}$ are the structure constants of $G$, while $k\in \bZ$ is a quantized coefficient that governs the 2-group structure. This term encodes the coupling of the photon to the current for a topological 0-form symmetry of the sigma model.
This result has a concrete application to the low-energy dynamics of 4d $U(N)$ QCD. For low enough number of flavors, the chiral symmetry is spontaneously broken and quarks form pion bound states as in $SU(N)$ QCD. However, here the theory also contains an Abelian gauge field $A$ for the baryon number symmetry with quarks charged under it, hence in the IR this photon cannot be decoupled. The photon-pion term encodes the coupling of $A$ to the baryon number current in the IR. We argue that the theory has a spontaneously-broken 2-group symmetry, implying that the leading photon-pion interaction coincides with the one we determined from our conjecture.

Since our work utilizes TQFTs with an infinite number of (simple) topological operators, as an aside in Appendix~\ref{app:TQFT} we explore some of their properties and show (in a simple example) that while their path integrals on closed Euclidean manifolds are divergent, the path integrals on open manifolds can be made finite.

The rest of the paper is organized as follows.
In Section~\ref{sec:TQFT holo} we explain the general setup and clarify some issues about holography with TQFTs in the bulk.
In the rest of the sections we present several interesting examples.
Section~\ref{sec: U(1) goldstone} concerns the vanilla example of Abelian symmetries without additional structures.
In Section~\ref{sec:Abelian anomalies} we include chiral anomalies and higher group structures, showing that the Goldstone theory is the same as in the vanilla case but it couples differently to background fields, a fact that is interpreted in terms of \emph{symmetry fractionalization}.
The non-invertible example is discussed in Section~\ref{sec:boundary chern-simons} after we warm up with a similar but simpler example in 3d that produces Maxwell--Chern--Simons theory.
The non-Abelian cases (including higher groups) are finally studied in Section~\ref{sec:non-Abelian}.

%%%%%%%%%%%%%%%%%%%%%%%%%%%%%%%%%%%%%%%%%%%%%%%%%%%%

\section{Topological field theories as holographic duals}
\label{sec:TQFT holo}

The bulk theories we use in this paper are TQFTs of the type introduced in \cite{Brennan:2024fgj, Antinucci:2024zjp, Bonetti:2024cjk} to describe SymTFTs for continuous symmetries. In the simplest cases, they have a Lagrangian formulation as%
\footnote{We only consider Euclidean manifolds and normalize our actions so that the weight in the path integral is $e^{-S}$.}
\be
\label{eq:continuousTFT}
S = \frac{i}{2\pi} \int_{X_{d+1}} b_{d-p-1} \wedge dA_{p+1}
\ee
where $A_{p+1}$ is a $U(1)$ $(p+1)$-form gauge field, while $b_{d-p-1}$ is an $\bR$ $(d-p-1)$-form gauge field. In the whole paper, we adopt this convention in which uppercase letters indicate $U(1)$ gauge fields, while lowercase letters indicate $\bR$ gauge fields. Understood as a SymTFT, this describes a $p$-form $U(1)$ symmetry in $d$ dimensions. The topological operators of the theory are \cite{Antinucci:2024zjp}:
\be
V_n(\gamma_{p+1}) = e^{in \int _{\gamma_{p+1}} A_{p+1}} \,, \quad U_\beta(\gamma _{d-p-1}) = e^{i\beta \int _{\gamma _{d-p-1}}b_{d-p-1}} \,, \quad n \in \bZ \,, \quad \beta \in \bR/\bZ \cong U(1) \,.
\ee
The partition function of \eqref{eq:continuousTFT} on a generic closed manifold diverges, but infinities are avoided on certain classes of manifolds with boundaries (see Appendix~\ref{app:TQFT}). These are the relevant ones for both the SymTFT and the holographic setup considered in this paper. Moreover, normalized correlators are always finite, and capture the braiding of topological defects:
\be
\big\langle V_n(\gamma_{p+1}) \, U_\beta(\gamma _{d-p-1}') \big\rangle = \exp \Bigl[ 2\pi i \, n \, \beta \operatorname{Link} \bigl( \gamma_{p+1} , \gamma _{d-p-1}' \bigr) \Bigr] \,.
\ee
In the following we will consider several modifications of the vanilla case \eqref{eq:continuousTFT} that take into account anomalies, higher groups, non-invertible symmetries, as well as extensions to non-Abelian groups. However let us focus here on this simplest case as an illustration of the basic ideas and setup.

\begin{figure}[t]
\centering
\begin{tikzpicture}[scale=1]
	\node at (2.5, 4) {\LARGE{SymTFT}};
	\draw [fill = red!60!black] (0,0) -- (1,1) -- (1,3) -- (0,2) node[left, align = left] {physical \\[-0.1em] QFT} -- cycle;
	\draw[black] (0,2)-- (4,2);
	\draw[black] (0,0)-- (4,0);
	\draw[black, dashed] (1,1)-- (5,1);
	\draw[black] (1,3)-- (5,3);
	\fill [fill = yellow, opacity = 0.2] (0,0) -- (4,0) -- (4,2) -- (5,3) -- (1,3) -- (0,2) -- cycle;
	\draw [fill = blue!30!white, opacity = 0.8] (4,0) -- (5,1) -- (5,3) -- (4,2) -- cycle;
	\node at (2.6, 1.5) {\Large{$\cZ(\cC)$}};
	\draw[black, ->, bend right=20] (5.5,-0.1) to (4.5,1.5); 
	\node[align = center] at (5.5,- 0.8) {topological \\[-0.2em] b.c. \large$\cL$}; 
\end{tikzpicture}
\hspace{1.5cm}
\begin{tikzpicture}[scale=1]
	\node at (1, 4.0) {\LARGE{Holography}};
	\draw [very thick, red!60!black] (0.8, 1.6) circle [radius = 1.6];
	\fill [fill=yellow, opacity=0.2] (0.8, 1.6) circle [radius = 1.6];
	\node at (0.8,1.6) {\Large$\dfrac{\cZ(\cC)}{\cL}$};
	\draw[black, ->, bend right=20] (3.7,-0.1) to (2.7,1.5); 
	\node[align=center] at (3.8, -0.8) {non-topological \\[-0.2em] b.c.}; 
\end{tikzpicture}
\caption{\label{fig: symmTFTvsholo}%
Left: the SymTFT setup. The TQFT is placed on a slab, whose right boundary is topological and determined by a Lagrangian algebra $\cL$.
Right: the holographic setup considered here. There is only one boundary with non-topological boundary conditions, while the Lagrangian algebra $\cL$ is gauged to make the bulk invertible.}
\end{figure}
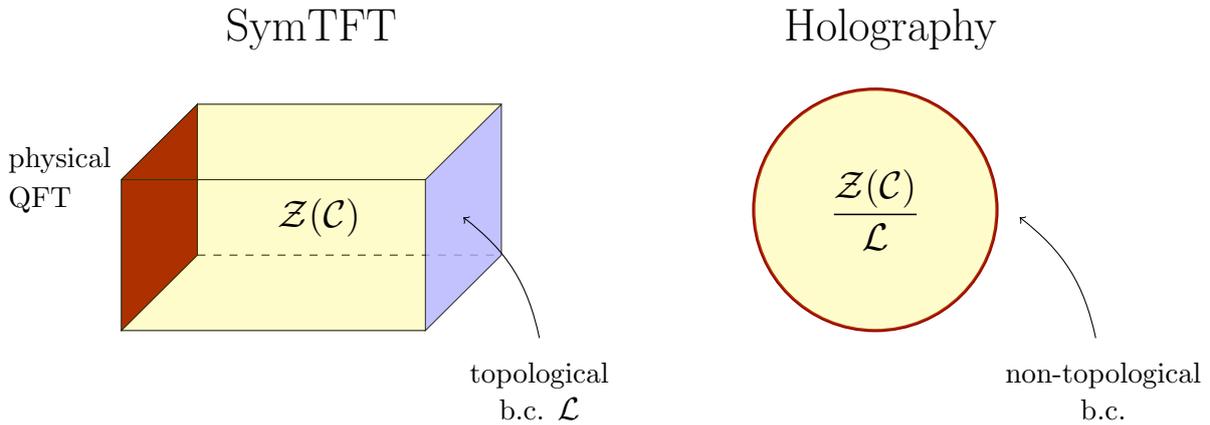

In SymTFT, \eqref{eq:continuousTFT} is placed on a slab with two boundaries, one of which is topological and determines the symmetry after the slab is squeezed. This topological boundary is characterized by a maximal set of mutually transparent objects, which we generically refer to as a Lagrangian algebra $\cL$. In this example a natural Lagrangian algebra consists of all $V_n(\gamma_{p+1})$, while the $U_\beta(\gamma _{d-p-1})$ become the generators of the $U(1)$ $p$-form symmetry of the boundary theory.  

In this paper, instead, we consider a different setting in which \eqref{eq:continuousTFT} is placed on a manifold $X_{d+1}$ with a unique connected boundary $\cM_d=\partial X_{d+1}$, which we endow with a Riemannian structure. On $\cM_d$ we fix non-topological boundary conditions$\,$%
\footnote{Such boundary conditions in BF theory, and the edge modes they lead to, have recently been studied in \cite{Fliss:2023uiv}.}
\be
\label{eq:bc sec2}
A_{p+1} + iC \, \star b_{d-p-1} = \cA_{p+1} \,.
\ee
Here $\star$ is the Hodge star operator of the boundary, $\cA_{p+1}$ is a fixed $(p+1)$-form on the boundary, and $C$ is a generically dimensionful constant with mass dimension $[C]=2p + 2 -d$.%
\footnote{The introduction of such a scale is necessary since the components of $A_{p+1}$ have dimension $p+1$ while those of $b_{d-p-1}$ have dimension $d-p-1$. In this way the forms $A_{p+1}$ and $b_{d-p-1}$ are dimensionless, the action in \eqref{eq:continuousTFT} is dimensionless, but $\star \, b_{d-p-1}$ has dimension $d-2p-2$.}
Moreover, the Lagrangian algebra $\cL$ that was used to define the topological boundary in the SymTFT setup must now be gauged in the bulk $X_{d+1}$, and the final bulk theory $\cZ(\cC)/\cL$ is an invertible TQFT. See Fig.~\ref{fig: symmTFTvsholo} for a comparison of the two setups.

In this second setup we want to establish a precise holographic duality with a certain local QFT$_d$ living on the boundary, which we need to determine. More precisely, the equality we need to show is the standard one \cite{Maldacena:1997re, Gubser:1998bc, Witten:1998qj}:
\be
\label{eq:holo}
Z _{\text{TQFT}_{d+1}} \bigl[ \varphi\big|_\partial =\cA \bigr] \,=\, Z_{\text{QFT}_d}[\cM_d, \cA] \,.
\ee
Here TQFT$_{d+1}$ is the result of gauging $\cL$ in $\cZ(\cC)$, $\varphi$ denotes generically some bulk fields (for instance $\varphi =A_{p+1} + iC \star b_{d-p-1}$ in the example \eqref{eq:continuousTFT}), while $\cA$ is introduced as a boundary value from the bulk viewpoint and plays the role of a background field for the boundary QFT.
Although SymTFT superficially resembles holography, the two are fundamentally different. SymTFT only captures symmetries and disregards dynamics, allowing any QFT with the specified symmetry. In contrast, in holography the dual QFT$_d$ is uniquely determined by the bulk theory and its boundary conditions, encoding both symmetries and dynamics as in \eqref{eq:holo}.%
\footnote{See \cite{Heckman:2024oot} for a general description of symmetry operators in holography.}

We will determine the dual QFT$_d$ explicitly in the many examples considered below, providing strong evidence for the conjecture that the dual theory to $\cZ(\cC)/\cL$ is always the symmetry-breaking EFT for $\cC$. Some of these checks are quite subtle and highly nontrivial. For instance, the Goldstone theory for a $U(1)$ symmetry with a cubic 't~Hooft anomaly in 4d is still a compact boson with no additional terms as in the non-anomalous case,%
\footnote{This is different from the non-Abelian case, in which an anomaly implies a WZW term in the sigma model.}
but the background field for the symmetry is coupled non-minimally to the theory. We discuss this in Section~\ref{sec:anomaly matching SSB} (in particular \eqref{eq:symm fract coupling} is the additional coupling) to which we refer for more details. The SymTFT for a 4d anomalous $U(1)$ is \cite{Antinucci:2024zjp}
\be
S = \frac{i}{2\pi} \int_{X_5} b_3 \wedge dA_1 + \frac{i k}{24 \pi^2} \int_{X_5} A_1\wedge dA_1 \wedge dA_1 \,.
\ee
Forgetting about the boundary value $\cA_1$ appearing in the boundary condition \eqref{eq:bc sec2}, the additional cubic term does not affect the dual boundary QFT$_4$. However we will show in Section~\ref{sec: Abelian anomaly 4d} that keeping track of $\cA_1$ we reproduce exactly the non-minimal coupling expected for an anomalous $U(1)$.

Before moving to the various examples, let us clarify a conceptual point. The assertion that certain dynamical QFTs have a TQFT as holographic dual might be perplexing at first. The origin of the confusion is that, even though TQFTs are good theories of gravity, the non-appearance of a metric tensor $g_{\mu\nu}$ is puzzling for holography: the metric should be dual to the stress-energy tensor $T_{\mu\nu}$ of the boundary QFT. While this observation is in general correct, in a few special cases it might have a loophole: the stress tensor might not be an independent operator. For instance, this is the case in the CS/WZW correspondence \cite{Witten:1988hf, Elitzur:1989nr}. In 2d WZW models the stress tensor of the CFT, using the Sugawara construction, is made out of the currents which are dual to the gauge fields of the 3d Chern--Simons bulk theory. Something very similar happens in our examples. Indeed, the EFTs for symmetry breaking are very special QFTs in which everything, including the stress-energy tensor, is determined by the currents and their correlation functions. This is at the core of the \emph{universality} of those EFTs. For instance, in the theory of a $U(1)$ Goldstone boson with action
\be
S = \frac{R^2}{4\pi}\int _{\cM_d}d\Phi \wedge \star \, d\Phi \,,
\ee
the $U(1)$ current is $J_\mu = \frac{iR^2}{2\pi}\partial _\mu \Phi$ and the stress tensor is a composite operator of $J_\mu$:
\be
T_{\mu\nu} = \frac{R^2}{4\pi} \biggl( \partial_\mu \Phi \, \partial_\nu \Phi - \frac{1}{2} \, \delta_{\mu\nu} \, (\partial \Phi)^2 \biggr) = \frac{\pi}{R^2} \, \biggl( \frac{1}{2} \, \delta_{\mu\nu} \, J^2 - J_\mu J_\nu \biggr) \,.
\ee
Through the boundary conditions, the bulk SymTFT provides background fields for the global symmetries of the boundary theory, which are sources for the boundary currents. Hence the TQFT can compute correlation functions of the currents, and by universality correlation functions of all operators, including those of the stress tensor, even without an explicit source $g_{\mu \nu}$. This is a general statement: in the EFTs for spontaneous breaking the currents completely determine all operators and the holographic duals do not need a graviton field.

It is expected, however, that embedding our models into RG flows and taking into account non-universal features would require to reintroduce dynamical gravity into the game. Indeed, a related observation is that the boundary theories we obtain are either free or non-renormalizable. The reason why a TQFT, which is expected to be UV complete and finite, can be dual to a non-renormalizable theory is the choice of non-topological boundary conditions, which introduce an energy scale in the theory. This scale sets a limit below which both the bulk and boundary theories are well defined. Above this threshold, the boundary theory requires the inclusion of more and more operators to tame UV divergencies. This issue has to carry over to the bulk TQFT as well --- albeit in a way unclear to us --- making the TQFT description incomplete. The expectation is that, to make sense of the bulk theory above the scale of the boundary condition, one has to allow for dynamical gravity in the bulk in a way that is similar to the embedding of an EFT for spontaneous breaking into a UV complete theory. It would be interesting to understand this point better.

%%%%%%%%%%%%%%%%%%%%%%%%%%%%%%%%%%%%%%%%%%%%%%%%%%%%

\section{\tps{\matht{U(1)}}{U(1)} Goldstone bosons}
\label{sec: U(1) goldstone}

The simplest cases to test our conjecture are those of $U(1)$ symmetries of generic degree. We warm up with the textbook example of a spontaneously broken $U(1)$ $0$-form symmetry in generic dimension and then move on to the case of higher-form symmetries, whose Goldstone bosons are (free) $U(1)$ higher-form gauge fields \cite{Gaiotto:2014kfa}.

\subsection{0-form symmetries}
\label{sec: 0-form goldstone}

Consider the following TQFT in $d+1$ dimensions:
\be
\label{action basic U(1) case}
S = \frac{i}{2\pi}\int_{X_{d+1}}b_{d-1}\wedge dA_1 \,,
\ee
where $A_1$ is a $U(1)$ gauge field while $b_{d-1}$ is an $\bR$ $(d-1)$-form gauge field. We endow the boundary $\cM_d = \partial X_{d+1}$ with a Riemmanian metric and impose the boundary condition
\be
\label{eq:bc in d dim}
\star \, A_1 = - \frac{i}{R^2} \, b_{d-1} + \star \, \cA_1 \,.
\ee
Here $R$ is a parameter of mass dimension $(d-2)/2$, while $\cA_1$ is a fixed background 1-form on $\cM_d$. Notice that only in $d=2$ this boundary condition is conformally invariant. In order to get a consistent variational principle with this boundary condition we must add a boundary term $S_\partial$ to (\ref{action basic U(1) case}). Indeed, the variation of the action produces a boundary piece
\be
\delta S \big|_{\cM_d} = (-1)^{d-1} \, \frac{i}{2\pi} \!\int _{\cM_d} \! b_{d-1}\wedge \delta A_1 = \frac{1}{2\pi R^2} \int_{\cM_d} b_{d-1} \wedge \star \, \delta b_{d-1} \,,
\ee
which requires a boundary term
\be
\label{eq:bdry term goldstone}
S_\partial = - \frac{1}{4\pi R^2 } \int_{\cM_d} b_{d-1}\wedge \star \, b_{d-1}  \,.
\ee
Since the boundary condition \eqref{eq:bc in d dim} breaks gauge invariance on the boundary, we have to be careful in specifying the group of transformations we quotient by in the bulk: we choose to allow only gauge transformations that are trivial on the boundary. This implies that the bulk gauge symmetries become global on the boundary. For any global symmetry we should be able to turn on a background. In our setup this operation has a very natural realization: instead of freezing gauge transformations on the boundary, we allow them but transform the boundary data so as to render the boundary condition invariant. For instance, we can make \eqref{eq:bc in d dim} gauge invariant under gauge transformations of $A_1$ by demanding that $A_1\mapsto A_1+d\lambda _0$ is accompanied by a transformation of the fixed background $\cA_1$:
\be
\cA_1 \mapsto \cA_1 + d\lambda _0 \,.
\ee
With this choice, $\cA_1$ is interpreted as a background gauge field for the global $U(1)$ symmetry on the boundary. Notice that with our choice of boundary term the whole system is gauge invariant.

We can also restore the gauge transformations $b_{d-1} \mapsto b_{d-1} + d\nu _{d-2}$ by transforming
\be
\cA_1 \,\mapsto\, \cA_1 - (-1)^d \frac{i}{R^2} \, \star d\nu _{d-2} \,,
\ee
which however are not proper background gauge transformations. A clearer and equivalent possibility is to parametrize the boundary condition as
\be
\label{eq:bcddimwind}
\star \, A_1 = - \frac{i}{R^2} \, \bigl (b_{d-1} - \cB_{d-1} \bigr) \,,
\ee
where $\cB_{d-1}$ is another fixed background on the boundary that transforms as $\cB_{d-1} \mapsto \cB_{d-1} + d\nu _{d-2}$. It can be understood as a background field for the global $(d-2)$-form symmetry on the boundary. Yet another possibility is to restore both gauge transformations, for instance through the parametrization
\be
\star \, \bigl( A_1 - \cA_1 \bigr) = - \frac{i}{R^2} \, \bigl( b_{d-1} - \cB_{d-1} \bigr) \,.
\ee
We can use it to discover information about the boundary theory. Indeed, with the choice of boundary term in (\ref{eq:bdry term goldstone}), the system is not gauge invariant, rather under a gauge transformation we find
\be
\label{eq:gauge variation compact boson}
\delta(S + S_\partial ) = (-1)^{d-1} \frac{i}{2\pi} \!\int_{\cM_d} \! d\nu _{d-2} \wedge \cA_1 - \frac{1}{4\pi R^2} \!\int_{\cM_d} \! \bigl( 2 \, d\nu_{d-2} \wedge \star \, \cB_{d-1} + d\nu_{d-2} \wedge \star \, d\nu_{d-2} \bigr) \,.
\ee
The second piece can be cancelled by modifying the boundary term with the addition of
\be
\frac{1}{4\pi R^2} \int_{\cM_d} \cB_{d-1}\wedge \star \, \cB_{d-1} \,,
\ee
that can be understood as a local counterterm. However the first piece in \eqref{eq:gauge variation compact boson} cannot be removed while preserving background gauge invariance for the $U(1)$ 0-form symmetry. This is a sign that the two symmetries have a mixed 't~Hooft anomaly. Indeed, as we are going to see, the theory we are describing is the holographic dual to a $d$-dimensional compact boson. In what follows we will turn on only the background for the $U(1)$ 0-form symmetry, \ie, we will use the boundary condition \eqref{eq:bc in d dim}.

In order to rewrite the path integral of this TQFT as that of the compact boson we proceed in analogy with \cite{Elitzur:1989nr, Moore:1989vd, Moore:1989yh} (see also \cite{Maldacena:2001ss}). We assume that $X_{d+1}$ contains an $S^{1}$ factor parametrized by $t\sim t+\beta$, interpreted as Euclidean time, hence $X_{d+1}= X_d \times S^1$ and $\partial X_{d+1} \equiv \cM_d = \cM_{d-1} \times S^1$. For simplicity, we also choose the metric of $ \partial X_{d+1}$ to be diagonal in $\cM_{d-1}$ and $S^1$ so that
\be
\star \, dt = (-1)^{d-1} \, \Vol_{\cM_{d-1}} \:\in\, \Omega^{d-1}(\cM_{d-1})
\ee
with $\Vol_{\cM_{d-1}}$ the volume form of $\cM_{d-1}$. We decompose the bulk fields as
\be
\label{eq:fields decompositions}
A_1 = A^t_0 \, dt + \widetilde{A}_1 \,, \qquad\qquad b_{d-1} = b^t_{d-2} \wedge dt+ \widetilde{b}_{d-1} \,,
\ee
where forms with a tilde live on the \emph{spatial} manifold $X_d$. The time components $A_0^t$ and $b_{d-2}^t$ appear linearly and can be treated as Lagrange multipliers. Integrating them out enforces
\be
\label{eq:spatial EOM}
\tilde{d} \widetilde{A}_1 = 0 \,, \qquad\qquad\qquad \tilde{d} \: \widetilde{b}_{d-1} = 0 \,.
\ee
We now make a choice for $X_{d}$ and take it to be a $d$-dimensional ball so that $\cM_d = S^{d-1}\times S^{1}$. Then \eqref{eq:spatial EOM} are solved by introducing a compact scalar $\Phi_0$ and a $(d-2)$-form $\bR$ gauge field $\omega_{d-2}$ as
\be
\label{eq:solution}
\widetilde{A}_1 = \tilde{d} \Phi_0 \,, \qquad\qquad\qquad \widetilde{b}_{d-1} = \tilde{d} \omega_{d-2} \,.
\ee
Rewriting both the bulk action and the boundary term using $\Phi_0$ and $\omega_{d-2}$, the system reduces to the boundary action
\bea
\label{eq:luttinger}
S &= \frac{i}{2\pi} \int_{\cM_d} \biggl[ (-1)^d \, \tilde{d} \omega _{d-2} \wedge \bigl( \partial_t \Phi _0 - \cA_0^t \bigr) dt \\
&\qquad\qquad\qquad - \frac{i}{2} \biggl( R^2 \bigl( \tilde{d} \Phi_0 - \widetilde{\cA}_1 \bigr) \wedge \star \, \bigl( \tilde{d} \Phi_0 - \widetilde{\cA}_1 \bigr) + \frac{1}{R^2} \, \tilde{d} \omega_{d-2} \wedge \star \, \tilde{d} \omega_{d-2} \biggr) \biggr] \,.
\eea
This action is not covariant, and time derivatives appear linearly. For $d=2$, the action contains two scalars and is a manifestly self-dual formulation of the compact boson known in the condensed matter literature as the Luttinger liquid Lagrangian (see, \eg, \cite{Cheng:2022sgb} for a recent discussion). It has the advantage of making both $U(1)$ symmetries explicit, at the expense of hiding Lorentz invariance. The action (\ref{eq:luttinger}) is a $d$-dimensional generalization of it and it makes both the 0-form and the $(d-2)$-form $U(1)$ symmetries manifest.

Path integrals with an action linear in time derivatives are interpreted as phase-space path integrals. One can typically obtain a configuration-space path integral by integrating out the momenta that appear quadratically. Indeed, here $\tilde{d} \omega_{d-2}$ is the conjugate momentum to $\Phi_0$ and we can recast the theory in a Lorentz-invariant form by integrating out $\omega_{d-2}$. An important observation is that the action has zero modes that need to be eliminated. One way to see this is via the equations of motion for $\omega_{d-2}$. These are
\be
\tilde{d} \, \biggl[ \bigl( \partial_t \Phi_0 - \cA_0^t \bigr) dt \,+\, (-1)^{d} \frac{i}{R^2} \star \tilde{d} \omega_{d-2} \biggr] = 0
\ee
with solution
\be
\tilde{d}\omega_{d-2} = i R^2 \, \bigl( \partial_t \Phi_0 - \cA_0^t \bigr) \star dt - i R^2 \star \tilde{d} \gamma_0 \,.
\ee
Notice that, since $\bigl( \partial_t \Phi_0 - \cA_0^t \bigr) \star dt$ is a $(d-1)$-form supported only on space, we have $\tilde{d} \star \tilde{d} \gamma_0=0$.
The scalar $\gamma_0$ is integrated over but its path integral is naively divergent because $\gamma_0$ has vanishing action, \ie, it is a zero-mode. Therefore in order to get a consistent theory we have to gauge fix $\gamma_0=0$. Plugging $\tilde{d}\omega _{d-2}$ in (\ref{eq:luttinger}) we get the final action
\be
\label{eq:goldstone U(1) action}
S = \frac{R^2}{4\pi}\int_{\cM_d} \bigl( d \Phi_0 - \cA_1 \bigr) \wedge \star \, \bigl( d \Phi_0 - \cA_1 \bigr) \,, 
\ee
corresponding to a $d$-dimensional compact boson with radius $R$. Had we integrated out $\Phi_0$ from \eqref{eq:luttinger}, we would have found the dual formulation in terms of the $(d-2)$-form $\omega_{d-2}$. The background field $\cA_1$ corresponds to the $U(1)$ shift symmetry of the boson and the anomalous shift we discussed above corresponds to the mixed 't~Hooft anomaly with the winding symmetry.

One might be puzzled by the fact that we have one bulk gauge symmetry $U(1)$, but we still obtain two global symmetries on the boundary, which might seem to clash with the usual holographic expectations. However, for the compact boson this is not really a contradiction: all correlation functions of one current can be obtained from those of the other. Indeed, the backgrounds of the two symmetries are obtained one from the other using the $\star$ operator (modulo counterterms, which correspond to contact terms in correlators); thus, functional derivatives of the partition function with respect to a single background already contain the information of all correlators of both currents (see \cite{Hofman:2017vwr} for a related discussion).

Before going on, let us mention an alternative, quicker way to arrive at the final result that does not pass through the Luttinger-liquid-like formulation \eqref{eq:luttinger}. It requires $X_{d+1}$ to be a ball, and hence $\cM_d=S^d$. After determining the boundary conditions \eqref{eq:bc in d dim} and the boundary term \eqref{eq:bdry term goldstone}, we just integrate the entire $b_{d-1}$ out, imposing $dA_1=0$. Since the bulk is now topologically trivial, this is solved by $A_1=d\Phi _0$. Using the boundary condition to express the boundary term \eqref{eq:bdry term goldstone} in terms of $A_1$, and plugging back $A_1=d\Phi _0$, we immediately get \eqref{eq:goldstone U(1) action}.

\subsection{Higher-form symmetries}

The higher-form case is very similar and we only flash the 1-form symmetry example, just to highlight one small subtlety. The TQFT we start with has action
\be
S = \frac{i}{2\pi} \int_{X_{d+1}} f_{d-2} \wedge dG_2 \,,
\ee
with $f_{d-2}$ and $G_2$ being an $\bR$ and $U(1)$ gauge field, respectively. On $X_{d+1}$ with boundary $\cM_d$, that we endow with a Riemannian metric (if $d=4$ a conformal structure is enough) we set the boundary condition (see also \cite{Fliss:2023uiv}):
\be
\label{eq:bc photon}
\star \, G_2 = (-1)^{d+1} \, \frac{i e^2}{\pi} \, f_{d-2} + \star \, \cG_ 2 \,,
\ee
where $[e^2]=4-d$. We must also add a boundary term
\be
S_\partial = - \frac{e^2}{4\pi^2} \int_{\partial X_{d+1}} f_{d-2} \wedge \star f_{d-2} \,.
\ee

When solving the constraints imposed by the integral over time components as 
\be
\widetilde{f}_{d-2} = \tilde{d}\omega_{d-3} \,, \qquad\qquad\qquad \widetilde{G}_2 = \tilde{d} A_1 \,,
\ee
we introduce (time-dependent) forms $\omega_{d-3}$ and $A_1$ only on the spatial manifold $X_d$, namely without time components. The boundary action one obtains is
\bea
S &=  \frac{i}{2\pi} \int_{\cM_d} \biggl[ (-1)^d \, \tilde{d} \omega_{d-3} \wedge \bigl( \partial_t A_1 + \cG_1^t \bigr) \wedge dt \\
&\qquad\qquad\qquad - \frac{i}{2} \biggl( \frac{e^2}{\pi} \, \tilde{d} \omega_{d-3} \wedge \star \, \tilde{d} \omega_{d-3} + \frac{\pi}{e^2} \, \bigl( \tilde{d} A_1 -\widetilde{\cG}_2 \bigr) \wedge \star \, \bigl( \tilde{d} A_1 - \widetilde{\cG}_2 \bigr) \biggr)  \biggr]  \,.
\eea
This is a higher-form generalization of \eqref{eq:luttinger} and integrating out $\omega _{d-3}$ we obtain
\be
S = \frac{1}{4 e^2} \int_{\cM_d} \bigl( dA_1 - \cB _2 \bigr) \wedge \star \, \bigl( dA_1 - \cB_2 \bigr) \,,
\ee
where $\cB_2 = - \cG_1^t \wedge dt + \widetilde{\cG}_2$ is a 2-form background field. This is a Maxwell action in $d$ dimensions coupled to a background field $\cB_2$ for its electric 1-form symmetry. 

The subtlety we want to point out is that $A_1$ does not have the time component, hence this is a gauge-fixed Maxwell action.%
\footnote{This subtlety does not arise in the quicker procedure described at the end of the last section.}
There is a gauge choice that arises naturally in this reduction procedure, that is, the temporal gauge. The same story goes through for any higher-form gauge field: the boundary action is always a generalized Maxwell theory in the temporal gauge (see \cite{Maldacena:2001ss} for a discussion on this point). It is important to keep this small subtlety in mind when looking at more complicated TQFTs that produce further interactions involving the photon. For instance, in Section~\ref{sec:boundary chern-simons} we will obtain Chern--Simons terms on the boundary, and we will have to keep in mind that they always arise in the temporal gauge.

\subsection{Lagrangian algebras and topological sectors}
\label{sec:Lagrangian algebras}

There is one very important caveat in the discussion of the previous two sections. Let us focus on the 0-form symmetry case for definiteness. We have shown that with the boundary condition we chose, the path integral of the TQFT can be rewritten as a path integral with the action of a compact boson \eqref{eq:goldstone U(1) action}. However, the domain is not the one of the physical theory. The reason is that when we solve \eqref{eq:spatial EOM} introducing $\Phi _0$ and $\omega_{d-2}$ as in \eqref{eq:solution}, these fields cannot wind around the time circle $S^1$. Hence what we established in Section~\ref{sec: 0-form goldstone} is that the TQFT partition function is equal to the zero-winding sector of a compact boson.%
\footnote{For $d=2$ the boundary spatial manifold is $S^1$, and since $\Phi_0$ is compact the path integral includes a sum over all windings around that spatial circle, but not around the time circle.}

However, it turns out that we can produce the path integral in \emph{any} fixed winding sector, simply by inserting a Wilson line $e^{in \int_{S^1} \! A_1}$ along the time circle in the bulk. The line pierces the spatial manifold $X_d$ at a point $P$, creating a nontrivial $(d-1)$-cycle $\Sigma_{d-1}\subset X_d$ and introducing a monodromy for $\widetilde{b}_{d-1}$ around it:
\be
\int_{\Sigma_{d-1}} \widetilde{b}_{d-1} = 2 \pi n \,.
\ee
To get the TQFT partition function with this insertion, consider a generator $\frac{\eta_{d-1}}{2\pi}$ of $H^{d-1}( X_d \smallsetminus P; \bZ)$, namely $\int _{\Sigma _{d-1}}\eta _{d-1}=2\pi$. The second equation in \eqref{eq:spatial EOM} is now solved by
\be
\label{sol in top sector}
\widetilde{b}_{d-1} = n \, \eta _{d-1} + \tilde{d} \omega_{d-2} \,.
\ee
With the same steps as before we obtain a path integral on boundary fields $\Phi_0$ and $\omega _{d-2}$, again over configurations of $\Phi_0$ with zero winding around the time circle, but with a modified action with respect to \eqref{eq:luttinger}:
\bea
\label{eq:luttinger winding}
S_n &=  \frac{i}{2\pi} \int_{\cM_d} \biggl[ (-1)^d \, \tilde{d} \omega_{d-2} \wedge \bigl( \partial_t \Phi_0 - \cA_0^t \bigr) dt \\
& \qquad\qquad\qquad - \frac{i}{2} \, \biggl( R^2 \, \bigl( \tilde{d} \Phi_0 - \widetilde{\cA}_1 \bigr) \wedge \star \, \bigl( \tilde{d} \Phi_0 - \widetilde{\cA}_1 \bigr) + \frac{1}{R^2} \, \tilde{d} \omega_{d-2} \wedge \star \, \tilde{d} \omega_{d-2} \biggr) \biggr] \\
&\quad - (-1)^d \, \frac{in}{2\pi} \int_{\cM_d} \cA_0^t \, \widehat{\eta}_{d-1} \wedge dt + \frac{n^2}{4\pi R^2} \int_{\cM_d} \widehat{\eta}_{d-1} \wedge \star \, \widehat{\eta}_{d-1} \,.
\eea
Here $\widehat{\eta}_{d-1}$ is the pull-back of $\eta_{d-1}$ on $\cM_d$. It is a top form on $\partial X_d \equiv \cM_{d-1}$ and one can make a choice for the representative $\eta_{d-1}$ in (\ref{sol in top sector}) such that $\widehat{\eta}_{d-1} = \frac{2\pi}{v} \, \Vol_{\cM_{d-1}}$ with $v = \int_{\cM_{d-1}} \! \Vol_{\cM_{d-1}}$ the volume of the boundary spatial slice. In particular $\star \, \widehat{\eta}_{d-1}= \frac{2\pi}{v} \, dt$. Plugging this back into \eqref{eq:luttinger winding} we obtain
\be
S_n = S_0 - in \theta + \frac{\pi \beta n^2}{v R^2} \qquad\qquad\text{where}\qquad \theta = (-1)^d \int _{S^1} \cA_0^t \, dt  \,.
\ee
Here $S_0$ is the action \eqref{eq:luttinger} written in terms of the periodic scalar in the Luttinger liquid form, which could be rewritten in the Lorentz covariant form \eqref{eq:goldstone U(1) action} that makes manifest its nature as a boson of radius $R$. Notice that $\theta \sim \theta + 2\pi$ has the interpretation of a chemical potential for the $U(1)$ 0-form symmetry. The partition function with the line inserted is then
\be
\label{eq:fixed n}
Z_n = Z_\text{pert} \, \exp \biggl( i n \theta - \frac{\pi \beta}{vR^2} \, n^2 \biggr)
\ee
where $Z_\text{pert}$ is the perturbative contribution due to a periodic boson.

We want to show our claim that, after we condense a Lagrangian algebra in the bulk, the partition function includes the sum over all topological sectors of the compact scalar, hence reproducing the physical partition function. The simplest Lagrangian algebra contains all the lines $W_n = e^{in \!\int\! A_1}$ and no surfaces $V_\alpha = e^{i \alpha \!\int b_{d-1}}$. Due to our choice of geometry, gauging this algebra is the same as summing over all lines inserted along the time circle, hence summing over all $n$ in \eqref{eq:fixed n}. The bulk interpretation of this sum is that we are computing the partition function of the SPT phase obtained by gauging the algebra, which we are taking as our theory of gravity. Hence using Poisson's summation formula we find%
\footnote{Here we are neglecting an extra factor $\sqrt{\beta/vR^2}$, since normalizations of the path integrals do not play a role in this paper. A similar factor is neglected in (\ref{top sectors 2nd eqn}).} 
\be
Z_\text{gravity} = \sum_{n \in \bZ} Z_n = Z_\text{pert} \sum_{w \in \bZ} \exp \biggl[ - \frac{\pi vR^2}{\beta} \Bigl( w + \frac{\theta}{2\pi} \Bigr)^2 \biggr] \,.
\ee
The right hand side is precisely the partition function of a compact boson of radius $R$ (with chemical potential $\theta$).

More generally, the bulk TQFT has other Lagrangian algebras consisting of the lines $W_{km}$ and the surfaces $V_{m'/k}$ for an integer number $k \in \bZ$. Condensing one of them produces a different SPT phase in the bulk, hence a different theory of gravity. In the SymTFT story this corresponds to gauging the $\bZ_k$ subgroups of the $U(1)$ symmetry at the boundary \cite{Antinucci:2024zjp}. Because of the chosen geometry, there are no $(d-1)$-cycles in the bulk and hence condensing this algebra simply means summing over all Wilson lines of charge multiple of $k$. The result is
\be
\label{top sectors 2nd eqn}
Z'_\text{gravity} = \sum_{m \in \bZ} Z_{km} = Z_\text{pert} \sum_{w \in \bZ} \exp \biggl[ - \frac{\pi v}{\beta} \, \Bigl( \frac{R}{k} \Bigr)^2 \Bigl( w + \frac{k\theta}{2\pi}\Bigr)^2 \biggr]
\ee
and the right-hand side can be interpreted as the partition function of a compact boson of radius $R'=R/k$. This is an orbifold of the previous boundary theory, which could be thought of as a different global form of the same theory.

We want to comment on a slightly different way to obtain a holographic dual to compact bosons, which also fits our proposal. We could have started with the TQFT of two $\bR$ gauge fields described by the action
\be
S = \frac{i}{2\pi} \int_{X_{d+1}} b_{d-1} \wedge d a_1 \,.
\ee
In this TQFT the charges of the Wilson lines $W_\alpha = e^{i\alpha \!\int\! a_1}$ are not quantized, and since there is no sum over fluxes,%
\footnote{An $\bR$ gauge field admits a gauge in which the connection is globally defined, therefore the field strength is an exact form and its integrals on compact submanifolds vanish.}
there is no identification among the charges of $V_\beta = e^{i\beta \!\int b_{d-1}}$. The spectrum of bulk operators is then larger, labelled by $\bR\times \bR$, and the corresponding braiding is the phase $e^{2\pi i \alpha \beta}$. Lagrangian algebras are classified by the choice of a real number $Q \in \bR_+$ and are given by \cite{Benini:2022hzx}
\be
\cL _Q = \bigl\{ W_{Qn} , V_{Q^{-1}m} \bigm| n,m \in \bZ \bigr\}  \,.
\ee
It was shown in \cite{Antinucci:2024zjp} that this TQFT is the SymTFT for two $U(1)$ symmetries, namely a 0-form and a $(d-2)$-form, with a mixed anomaly. While this is a different symmetry structure from just a single $U(1)$, the second higher-form symmetry arises universally in the IR whenever the 0-form symmetry is spontaneously broken. Hence the two symmetry structures share the same EFT that describes the broken phase and, according to our proposal, they should both be the holographic dual to a compact boson. Indeed there is no much difference between the two theories: the non-topological boundary conditions can be chosen to be the same, and the computations of Section~\ref{sec: 0-form goldstone} give the same result. 

The considerations explained in this section can be repeated for any higher-form symmetry. However, in order to detect the various global structures of a boundary $p$-form Maxwell theory, one needs to properly choose the geometry. Indeed the fluxes are supported on $(p+1)$-dimensional cycles, and thus a natural choice is to take $X_{d+1} = B_{d-p} \times T^{p+1}$ with $B_{d-p}$ a ball. One of the $S^1$ factors of the torus plays the role of a time circle, and $X_d = B_{d-p} \times T^p$. The bulk TQFT has action
\be
S = \frac{i}{2\pi} \int_{X_{d+1}} b_{d-p-1} \wedge dA_{p+1}
\ee
where $b_{d-p-1}$ is an $\bR$ gauge field whilst $A_{p+1}$ is a $U(1)$ gauge field. One can obtain an SPT phase by gauging the Lagrangian algebra given by $W_n = e^{i n \!\int\! A_{p+1}}$, and this is realized by inserting these defects along the $T^{p+1}$ factor in the bulk. This sum indeed reproduces the sum over fluxes of the $p$-form Maxwell theory on the boundary. The choice of other Lagrangian algebras modifies the value of the electric charge and corresponds to discrete gaugings of the 1-form symmetry.

%%%%%%%%%%%%%%%%%%%%%%%%%%%%%%%%%%%%%%%%%%%%%%%%%%%%

\section{Abelian anomalies and higher groups}
\label{sec:Abelian anomalies}

We can enrich the analysis of $U(1)$ symmetries by including anomalies (Sections~\ref{sec: Abelian anomaly 2d} and \ref{sec: Abelian anomaly 4d}) or a 2-group structure (Section~\ref{sec: 2-groups}). We show here that, when doing it, the dual boundary theory gets coupled to background fields in a non-minimal way. In Sections~\ref{sec:anomaly matching SSB} and \ref{sec:2-group SSB} we provide a field-theoretic interpretation of our results in terms of symmetry fractionalization.

\subsection{Chiral anomaly in 2d}
\label{sec: Abelian anomaly 2d}

The SymTFT for an anomalous $U(1)$ symmetry in 2d has action \cite{Antinucci:2024zjp}:
\be
S = \frac{i}{2\pi} \int_{X_3} b_1\wedge dA_1 + \frac{i k}{4\pi} \int_{X_3} A_1 \wedge dA_1 \,.
\ee
The additional bulk Chern--Simons term significantly affects the consistent boundary conditions. To establish a proper variational principle with a non-topological boundary condition, it is essential to include the boundary term
\be
S_\partial = - \frac{1}{4\pi R^2} \int_{\partial X_3} \biggl(b_1 + \frac{k}{2} \, A_1 \biggr) \wedge \star \, \biggl( b_1 + \frac{k}{2} \, A_1 \biggr) 
\ee
together with the following Dirichlet boundary condition:%
\footnote{One can check, by writing all possible boundary terms and imposing consistency of the variational principle, that these boundary data are the only possible choice.}
\be
\label{eq:variation anomaly}
\star \, \delta A_1 = - \frac{i}{R^2} \, \delta\biggl( b_1 + \frac{k}{2} \, A_1 \biggr) \,.
\ee
In order to properly turn on a background for the boundary $U(1)$ symmetry we have to render the boundary condition invariant under gauge transformations of $A_1$. This is most naturally done by introducing a 1-form $\cA_1$ as
\be
\star \,  (A_1 - \cA_1) = - \frac{i}{R^2} \, \biggl( b_1 + \frac{k}{2} \, (A_1 - \cA_1) \biggr) \,.
\ee
This boundary condition is invariant under $\delta \cA_1 = \delta A_1 = d \lambda_0$, allowing us to interpret $\cA_1$ as a background field for the $U(1)$ symmetry on the boundary. Notice that our choice does not modify \eqref{eq:variation anomaly} and is thus just a particularly convenient parametrization.

Before deriving the dual boundary theory, we can already establish that it has an 't~Hooft anomaly. Indeed, under a gauge transformation $\delta A_1 = \delta \cA_1 = d\lambda_0$ the total action $S+S_\partial$ transforms as
\be
\delta (S + S_\partial) = - \frac{ik}{4\pi} \int_{\cM_2} d \lambda _0 \wedge \cA_1 - \frac{k^2}{16\pi R^2} \int_{\cM_2} \Bigl( 2 \, d \lambda _0 \wedge \star \, \cA_1 + d \lambda_0 \wedge \star \, d\lambda_0 \Bigr)
\ee
where $\cM_2 = \partial X_3$.
The second term can be cancelled by adding the following counterterm to the boundary action:
\be
\label{eq:counterterm}
S_{\text{c.t.}} = \frac{k^2}{16\pi R^2} \int_{\cM_2} \cA_1 \wedge \star \, \cA_1 \,.
\ee
However the remaining total gauge variation
\be
\label{eq:anomalous variation 2d}
\delta \bigl( S + S_\partial + S_{\text{c.t.}} \bigr) = - \frac{ik}{4\pi}\int _{\cM_2} d \lambda _0 \wedge  \cA_1 
\ee
cannot be cancelled by any local boundary counterterm: it is precisely the anomalous variation corresponding to a perturbative $U(1)$ anomaly. 

To derive the boundary theory we follow the steps outlined in Section~\ref{sec: U(1) goldstone}. The constraints imposed by the path integral over time components again allow us to write $\widetilde{A}_1 = \tilde{d} \Phi_0$ and $\widetilde{b}_1 = \tilde{d}\omega_0$. The boundary action expressed in terms of these variables, after introducing $\cF = \cA_1 - \frac{i k}{2 R^2} \star \cA_1$ for convenience, reads:
\begin{align}
S &= \frac{i}{2\pi} \int_{\cM_2} \biggl[ \Bigl( \tilde{d} \omega_0 + \tfrac{k}{2} \, \tilde{d} \Phi_0 \Bigr) \bigl( \partial_t \Phi_0 - \cF_0^t \bigr) \wedge dt \\
&\qquad - \frac{i}{2} \biggl( R^2 \Bigl( \tilde{d} \Phi_0 - \widetilde{\cF}_1 \Bigr) \wedge \star \, \Bigl( \tilde{d} \Phi_0 - \widetilde{\cF}_1 \Bigr) + \frac{1}{R^2} \Bigl( \tilde{d}\omega_0 + \tfrac{k}{2} \, \tilde{d} \Phi_0 \Bigr) \wedge \star \, \Bigl( \tilde{d} \omega_0 + \tfrac{k}{2} \, \tilde{d} \Phi_0 \Bigr) \biggr) \biggr] + S_\text{c.t} \,. \nn
\end{align}
This is the same action as in \eqref{eq:luttinger} for $d=2$ but with $\omega_0 \mapsto \omega_0 + \frac{k}{2} \, \Phi_0$. Integrating $\omega_0$ out we find
\be
\label{eq:2d U(1) anomaly lagrangian}
S= \frac{R^2}{4\pi} \int_{\cM_2} \bigl( d \Phi_0 - \cA_1 \bigr) \wedge \star \, \bigl( d\Phi_0 - \cA_1 \bigr) + \frac{ik}{4\pi} \int_{\cM_2} \Phi_0 \, d\cA_1 \,.
\ee
This action describes a compact boson of radius $R$, but with an unusual coupling to a background for the momentum symmetry. Such a coupling reproduces the anomalous shift \eqref{eq:anomalous variation 2d} that is indeed cancelled by the inflow action
\be
S_\text{inflow} = - \frac{ik}{4\pi} \int_\text{3d} \cA_1 \wedge d\cA_1 \,.
\ee

Notice that the extra coupling $\Phi _0 \, d\cA_1$ in \eqref{eq:2d U(1) anomaly lagrangian} has a form similar to the coupling with the winding symmetry. In a sense, we are prescribing that a background $\cA_1$ for the momentum symmetry also activates a background $\cB_1 = k\cA_1$ for the winding symmetry. In other words, $\cA_1$ is not coupled with the momentum symmetry but rather with a diagonal combination of momentum and winding.%
\footnote{More precisely, it is the diagonal combination between momentum and a $\bZ_k$ extension of the winding symmetry.}
Since the two symmetries have a mixed anomaly, this diagonal $U(1)$ inherits a pure anomaly.

\subsection{Chiral anomaly in 4d}
\label{sec: Abelian anomaly 4d}

The treatment of anomalies in higher dimensions presents a further conceptual difference. As a representative case, we consider $d=4$ and the TQFT with action
\be
S = \frac{i}{2\pi} \int_{X_5} b_3 \wedge dA_1 + \frac{i k}{24 \pi^2} \int_{X_5} A_1 \wedge dA_1 \wedge dA_1 \,.
\ee
To get a good variational principle we need to impose
\be
\star \, \delta A_1 = -\frac{i}{R^2} \, \delta \biggl( b_3 + \frac{k}{6\pi} A_1 \wedge dA_1\biggr)
\ee
and add a boundary term 
\be
S_\partial = - \frac{1}{4\pi R^2} \int_{\partial X_5} \biggl( b_3 + \frac{k}{6\pi} A_1 \wedge dA_1 \biggr) \wedge \star \, \biggl( b_3 + \frac{k}{6\pi} A_1 \wedge dA_1 \biggr) \,.
\ee
These choices however do not allow us to turn on a background by simply changing the parametrization of the boundary condition, as we did in 2d. Indeed, if we try to restore the gauge transformations of $A_1$, the boundary condition shifts by terms that depend on the field $A_1$ itself and cannot be cancelled by adding counterterms in the background only. Turning on a background in $d>2$ requires us to change the boundary data in a nontrivial way. In Appendix~\ref{app:anomalousbc} we explain an iterative procedure that, starting from the data above, produces a consistent variational principle together with a gauge-invariant boundary condition. The result for $d=4$ is
\be
\star \, \bigl( A_1 - \cA_1 \bigr) = - \frac{i}{R^2} \biggl( b_3 + \frac{k}{6\pi} \bigl( A_1 - \cA_1 \bigr) \wedge dA_1 + \frac{k}{12\pi} \bigl( A_1 - \cA_1 \bigr) \wedge d\cA_1\biggr)
\ee
with boundary term%
\footnote{We used the shorthand notation $\omega^2 \,\equiv\, \omega \wedge \star \, \omega$.}
\be
S_\partial = - \frac{1}{4\pi R^2} \!\int_{\partial X_5} \biggl( b_3 + \frac{k}{6\pi} \bigl( A_1 - \cA_1 \bigr) \wedge dA_1 + \frac{k}{12\pi} A_1 \wedge d\cA_1 \biggr)^{\!\!2} +\frac{ik}{24\pi^2} \!\int_{\partial X_5} \cA_1 \wedge A_1 \wedge dA_1 \,.
\ee
When setting $\cA_1 =0$ we recover the previous boundary data, but in general there are new terms that mix background and dynamical fields. As in 2d, one can show that the system has an anomaly performing a gauge transformation $\delta A_1 = \delta \cA_1 = d \lambda_0$: up to a counterterm the gauge variation is 
\be
\label{eq:anomalousshift4d}
\delta \bigl( S + S_\partial +S_\text{c.t.} \bigr) = \frac{ik}{24\pi ^2} \int_{\partial X_5} \lambda_0 \,  d\cA_1 \wedge d\cA_1 \,.
\ee

The procedure to determine the dual boundary theory is completely analogous to the examples we have already presented. Integrating the time components out, we introduce $\widetilde{A}_1 = \tilde{d} \Phi_0$ and $\widetilde{b}_3 = \tilde{d} \omega_2$. To simplify our expressions, we denote $\cF_1 = \cA_1 - \frac{i k}{12\pi R^2} \star (\cA_1 \wedge d\cA_1)$. Then the boundary action, in its non-covariant presentation, is
\begin{align}
S &= \frac{i}{2\pi} \int_{\cM_4} \biggl[ \Bigl( \tilde{d} \omega_2 + \tfrac{k}{12\pi} \, \tilde{d} \Phi_0 \wedge \tilde{d} \widetilde{\cA}_1 \Bigr) \bigl( \partial_t \Phi_0 - \cF_0^t \bigr) dt - \frac{i}{2} \biggl( R^2 \Bigl( \tilde{d} \Phi_0 - \widetilde{\cF}_1 \Bigr) \wedge \star \, \Bigl( \tilde{d} \Phi_0 - \widetilde{\cF}_1 \Bigr)  \nn \\
&\qquad\qquad\qquad\qquad + \frac{1}{R^2} \Bigl( \tilde{d} \omega_2 + \tfrac{k}{12\pi} \, \tilde{d} \Phi_0 \wedge \tilde{d} \widetilde{\cA}_1 \Bigr) \wedge \star \, \Bigl( \tilde{d} \omega_2 +\tfrac{k}{12\pi} \, \tilde{d} \Phi_0 \wedge \tilde{d} \widetilde{\cA}_1 \Bigr) \biggr) \biggr] + S_\text{c.t.}
\end{align}
where $\cM_4 = \partial X_5$. As before we can integrate out $\omega _2$ and the final action reads
\be
S = \frac{R^2}{4\pi} \int_{\cM_4} \bigl( d\Phi_0 - \cA_1 \bigr) \wedge \star \, \bigl( d\Phi_0 - \cA_1 \bigr) + \frac{i k}{24\pi^2} \int_{\cM_4} \Phi_0 \, d\cA_1 \wedge d\cA_1 \,.
\ee
This represents a compact scalar with a non-standard coupling to a background associated with the shift symmetry, akin to the situation in 2d. The additional interaction accounts for the anomalous variation described by (\ref{eq:anomalousshift4d}). Nevertheless, unlike in the 2d scenario, we cannot view this altered interaction as a combination of the shift and winding symmetries since the two have different degree.

\subsection{Anomaly matching in the broken phase}
\label{sec:anomaly matching SSB}

Let us provide a purely field-theoretic interpretation of the result in the previous section. For any Lie-group symmetry $G$, the Goldstone theory describing the symmetry breaking phase is a non-linear sigma model with target space $G$. In even spacetime dimensions $d$, the symmetry $G$ can suffer from perturbative anomalies and the question is how these are matched in the sigma model. 

For non-Abelian $G$ it is well known that the anomaly is reproduced by a WZW term \cite{Witten:1983tw}. This is an additional interaction with important dynamical consequences. Perturbative anomalies are classified by $H^{d+2}(BG;\bZ)$, which determines a $(d+1)$-dimensional Chern--Simons action that cancels the anomaly by inflow. On the other hand, WZW terms in $d$ dimensions are classified by $H^{d+1}(G;\bZ)$. Anomaly matching is mathematically represented by a map
\be
\tau : H^{d+2}(BG; \bZ) \rightarrow H^{d+1}(G; \bZ)
\ee
called \emph{transgression} \cite{Borel1955TopologyOL}. For $d=2$ this map also underlines the map of levels in the CS/WZW correspondence \cite{Dijkgraaf:1989pz}. For the simple Lie group $G=SU(n)$, the transgression map $\tau$ is injective \cite{Dijkgraaf:1989pz}, meaning that any perturbative anomaly is matched by a WZW term.%
\footnote{The transgression map is expected to be injective for all simple Lie groups.}
However this is not the general case, and if $\tau$ has a nontrivial kernel, the corresponding anomalies require some new ingredient to be matched in the sigma model.

Here we focus on the extreme case $G=U(1)$ for which $H^{d+1} \bigl( U(1); \bZ \bigr) = 0$, namely there is no WZW term at all, and any anomaly must be matched in a different way. From our holographic analysis we know the answer to this question: the dynamics of the sigma model is unchanged with respect to the non-anomalous case, but the symmetry is coupled non-minimally to the background $\cA_1$ through the extra topological term
\be
\label{eq:symm fract coupling}
\frac{ik}{(2\pi)^{d/2} \, \bigl( \frac{d}{2} + 1 \bigr)!} \, \int_{\cM _d} \Phi _0 \, \bigl( d \cA_1 \bigr)^{d/2} \,.
\ee
This term reproduces the anomaly, but at this level it seems a bit ad hoc. We want to clarify why it arises from a UV viewpoint and how we understand it in the IR. This is important to understand why there is a difference in how anomaly matching works in the Abelian and non-Abelian cases.

We can show in a simple model that when the background field is turned on in the UV, the additional coupling \eqref{eq:symm fract coupling} is generated along the RG flow by integrating out massive fields. Consider a 4d theory with a massless Dirac fermion $\psi$ and a complex scalar $\phi$, coupled via a Yukawa interaction:
\be
\cL \,\supset\, \phi \, \overline{\psi}\psi \,.
\ee
The theory has an axial symmetry $U(1)_A$ under which both Weyl components of $\psi$ have charge $1$, while $\phi$ has charge $-2$. $U(1)_A$ has a cubic anomaly with $k=2$. Choosing a potential $V(\phi)$ that induces condensation of $\phi$, the axial symmetry gets spontaneously broken to $\bZ_2 = (-1)^F$. By decomposing $\phi = \rho \, e^{i\Theta}$ into its radial and angular parts, the VEV $\langle \rho \rangle =v$ gives mass to both $\rho$ and $\psi$. The angular part $\Theta$ remains massless and is the only degree of freedom at low energy: it is the Goldstone boson. The faithful symmetry in the IR is $U(1)=U(1)_A/\bZ_2$ that shifts $\Theta$. In order to reproduce the anomaly, the coupling to a background $\cA$ must include the term
\be
\label{eq:coupling example}
\frac{i}{24\pi ^2} \, \Theta \, (d\cA)^2 \,.
\ee
Indeed this term arises when integrating out the fermion. To see this notice that, for fixed $\phi$ and $\cA$, if $\phi$ is real and positive then the fermion path integral can be regularized in a way such that the measure is positive  \cite{Weingarten:1983uj, Vafa:1983tf, Vafa:1984xg}. Clearly this is not true on a generic configuration, but we can make it true by performing an axial rotation of parameter $e^{i\alpha}$, with $\alpha = - \frac{1}{2} \Theta$. A textbook computation \cite{Fujikawa:1979ay, WeinbergQFTbook2} shows that the path integral measure of the fermion changes by a phase
\be
D[\psi] \,\mapsto\, D[\psi] \, \exp \biggl( \frac{ik}{24\pi^2} \int\! \alpha \, (d\cA)^2 \biggr) \,.
\ee
Setting $\alpha =-\frac{1}{2}\Theta$ this precisely reproduces the coupling \eqref{eq:coupling example}. Now the Yukawa coupling becomes $\rho \, \overline{\psi} \psi$, that for fixed $\rho$ is essentially a positive mass term for the fermion, hence integrating out the fermion becomes a safe operation that does not introduce extra phases.

Returning to the general case, we want to interpret the extra coupling \eqref{eq:symm fract coupling} as specifying a (higher) symmetry fractionalization class for the $U(1)$ symmetry. This reinterpretation will be crucial to understand the analogous story for higher groups in the following sections. A 0-form symmetry $G$ can fractionalize in the presence of a discrete 1-form symmetry $\Gamma$. This means that when two topological defects $g,h\in G$ fuse to produce $gh\in G$, their codimension-two junction gets covered by a topological defect $\omega (g,h) \in \Gamma$ of the 1-form symmetry \cite{Benini:2018reh, Delmastro:2022pfo, Brennan:2022tyl}, where $\omega \in H^2(BG; \Gamma)$. Equivalently, a background $\cA_1$ for $G$ turns on a background $\cB_2 = \cA_1^* \, \omega$ for the 1-form symmetry. In this formula, we think of $\cA_1$ as a map $\cM_d \rightarrow BG$ and of $\cB_2$ as an element of $H^2(\cM_d, \Gamma)$ so that we can use $\cA_1$ to pull back $\omega$. With this interpretation it becomes clear that, if $G$ and $\Gamma$ have a mixed anomaly, a non-trivial fractionalization class modifies the pure anomaly for $G$, possibly making it nontrivial even when it vanished originally \cite{Delmastro:2022pfo, Brennan:2022tyl}. This has a natural generalization to the case that $\Gamma$ is a discrete $p$-form symmetry: when $p+1$ topological defects $g_1, \dots ,g_{p+1} \in G$ fuse in generic position, they create a codimension-$(p+1)$ junction that can be dressed by a defect $\omega (g_1, \ldots,g_{p+1})$ of the $p$-form symmetry $\Gamma$, where $\omega$ is a class in $H^{p+1}(BG; \Gamma)$. Equivalently, a background $\cA_1$ turns on a background $\cB_{p+1} = \cA_1^* \, \omega$ for $\Gamma$.

The compact boson theory that describes the breaking of a $U(1)$ 0-form symmetry also possesses a $U(1)$ $(d-2)$-form winding symmetry, and the two have a mixed anomaly. For this reason, a pure anomaly for the 0-form symmetry can be induced by fractionalizing it with the $(d-2)$-form symmetry. One minor modification with respect to what we described above is necessary because the $p$-form symmetry (here $p=d-2$) is continuous. Its most natural description is not in terms of a background potential $\cB_{p+1}$, which is not a cohomology class in general, but in terms of its field strength $\frac1{2\pi} \, d\cB_{p+1} \in H^{p+2}(\cM_d; \bZ)$. As a consequence the fractionalization class, instead of being an element of $H^{p+1} \bigl( BU(1); U(1) \bigr)$, is more naturally an element of $H^{p+2}(BU(1); \bZ)\cong \bZ$. This is the datum that determines a $(p+1)$-dimensional Chern--Simons level, or equivalently the corresponding Chern class in $(p+2)$ dimensions. Hence, in analogy with the discrete case, we prescribe that a background $\cA_1$ for the 0-form symmetry activates a background $\cB_{d-1}$ for the $(d-2)$-form symmetry whose field strength is 
\be
\frac1{2\pi} \, d\cB_{d-1} = \frac{k}{(2\pi )^{d/2} \, \bigl( \frac{d}{2}+1 \bigr)!} \, \bigl( d\cA_1 \bigr)^{d/2} \,.
\ee
Recalling that the $(d-2)$-form symmetry is coupled to its background field through the action term $\frac{i}{2\pi} \!\int_{\cM_d} \! \Phi _0 \, d\cB_{d-1}$, this reproduces the coupling \eqref{eq:symm fract coupling} in agreement with our holographic result.

\subsection{Abelian 2-groups}
\label{sec: 2-groups}

We consider a 2-group symmetry in four dimensions formed by a $U(1)$ 0-form symmetry and a $U(1)$ 1\nobreakdash-form symmetry. This can be obtained by starting from a theory with two $U(1)$ 0-form symmetries with a cubic mixed anomaly and gauging the $U(1)$ that appears linearly in the anomaly polynomial \cite{Tachikawa:2017gyf, Cordova:2018cvg}. The $1$-form symmetry participating in the $2$-group structure is the magnetic symmetry of the photon. The SymTFT for such a 2-group symmetry has action \cite{Antinucci:2024zjp}:
\be
\label{eq:symTFT 2-group}
S = \frac{i}{2\pi} \int_{X_5} \biggl( b_3 \wedge dA_1 + h_2 \wedge dC_2 + \frac{k}{2\pi} \, h_2 \wedge A_1 \wedge dA_1 \biggr) \,. 
\ee
Here $A_1$ and $C_2$ are $U(1)$ gauge fields, while $b_3$ and $h_2$ are $\bR$ gauge fields. The topological operators that implement the symmetry are the Wilson surfaces of $b_3$ and $h_2$. On the other hand, the endpoints of $e^{i \!\int\! A_1}$ are local operators charged under the 0-form symmetry, and the endlines of $e^{i \!\int\! C_2}$ are 't~Hooft lines charged under the magnetic 1-form symmetry. The gauge transformations are:%
\footnote{There is some freedom in the choice of transformations that leave (\ref{eq:symTFT 2-group}) invariant. In particular, the transformation $\delta A_1 = d\lambda_0$ could be accompanied by an action on both $b_3$ and $C_2$ as $\delta b_3 = - \epsilon \frac{k}{2\pi} d\lambda_0 \wedge h_2$ and $\delta C_2 = (1-\epsilon) \frac{k}{2\pi} d\lambda _0 \wedge A_1$ for any choice of $\epsilon$. Here we chose $\epsilon=0$ which matches the transformations in the boundary theory.}
\be
\delta A_1 = d\lambda_0 \,,\qquad \delta h_2 = d\xi_1 \,,\qquad \delta b_3 = d\gamma_2 - \frac{k}{2\pi} \, d\xi_1 \wedge A_1 \,,\qquad \delta C_2 = d\eta_1 + \frac{k}{2\pi} \, d\lambda_0 \wedge A_1  \,.
\ee

We place this TQFT on a manifold with boundary,  $X_5 = B_4 \times S^1$ for simplicity, and we interpret it as a theory of gravity, holographically dual to some 4d quantum field theory on the boundary. The last term in \eqref{eq:symTFT 2-group} contains a derivative, therefore it affects the boundary contribution to the variational principle, similarly to the case of chiral anomalies. To fix the boundary terms $S_\partial$ and the boundary conditions on the fields, we use the same logic as in that case. We find  the boundary conditions
\be
\star \, \bigl( A_1 - \cA_1 \bigr) = - \frac{i}{R^2} \biggl[ b_3 + \frac{k}{2\pi} \, h_2 \wedge \bigl( A_1 - \cA_1 \bigr) \biggr] ,\quad \star \, h_2 = \frac{ie^2}{\pi} \biggl( C_2 - \cC_2 -\frac{k}{2\pi} \, \cA_1 \wedge A_1 \biggr)
\ee
and a corresponding boundary term 
\begin{align}
S_\partial &= - \frac{i}{2\pi} \!\int_{\partial X_5} \! h_2 \wedge \Bigl( C_2 - \tfrac{k}{2\pi} \, \cA_1 \wedge A_1 \Bigr) - \frac{e^2}{4\pi^2} \!\int_{\partial X_5} \! \Bigl( C_2 - \tfrac{k}{2\pi} \, \cA_1 \wedge A_1 \Bigr) \wedge \star \, \Bigl( C_2 - \tfrac{k}{2\pi} \, \cA_1 \wedge A_1 \Bigr) \nn \\
&\quad\, - \frac{1}{4\pi R^2} \!\int_{\partial X_5} \biggl[ b_3 + \frac{k}{2\pi} \, h_2 \wedge \bigl( A_1 - \cA_1 \bigr) \biggr] \wedge \star \, \biggl[ b_3 + \frac{k}{2\pi} \, h_2 \wedge \bigl( A_1 - \cA_1 \bigr) \biggr] . 
\end{align}
Here $\cA_1$, $\cC_2$ are fixed gauge fields on the boundary that transform as a proper 2-group background:
\be
\delta \cA_1 = d\lambda_0 \,, \qquad\qquad \delta \cC_2 = d\eta _1 + \frac{k}{2\pi} \, d\lambda_0 \wedge \cA_1 \,.
\ee
This makes the boundary conditions gauge invariant, provided we add a counterterm $\frac{e^2}{4\pi^2} \!\int_{\partial X_5} \cC_2 \wedge \star \, \cC_2$.

With the usual procedure, we obtain that the dual boundary theory has action:
\bea
\label{eq:SSB Abelian 2group}
S &= \frac{R^2}{4\pi} \int_{\partial X_5} \bigl( d \Phi_0 - \cA_1 \bigr) \wedge \star \, \bigl( d \Phi_0 - \cA_1 \bigr) + \frac{1}{4e ^2} \int_{\partial X_5} d a_1 \wedge 
\star \, d a_1 \\
&\quad + \frac{i}{2\pi} \int_{\partial X_5} \cC_2 \wedge d a_1 + \frac{ik}{4\pi^2} \int_{\partial X_5} \Phi_0 \; da_1 \wedge d\cA_1 \,.
\eea
Naively one may think that $a_1$ is an $\bR$ gauge field, because it comes from the trivialization of $h_2$. However, we have to take into account the condensation of the appropriate Lagrangian algebra in the bulk, necessary to trivialize the TQFT and making it independent of the topology. Specifically, here the relevant Lagrangian algebra is
\be
\cL = \Bigl\{  e^{in \!\int\! A_1} ,\, e^{im \!\int\! C_2} \Bigm| n,m \in \bZ \Bigr\} \,.
\ee
Following the same logic as in Section~\ref{sec:Lagrangian algebras}, this introduces a sum over the fluxes of $da_1$ that effectively makes $a_1$ into a $U(1)$ gauge field.

Turning off the background $\cA_1$ we obtain a free compact scalar and a free photon (coupled to a background field $\cC_2$ for its magnetic symmetry), enjoying a $U(1)$ 0-form symmetry with conserved current $J_1=\frac{iR^2}{2\pi} \, d\Phi_0$, and a $U(1)$ 1-form symmetry with conserved current $J_2 = \frac{1}{2\pi}\star da_1$, respectively. However, as soon as we turn on a background $\cA_1$ for the 0-form symmetry, the 2-group structure manifests itself through the nonstandard coupling between the photon and the scalar, which modifies the currents and the background gauge transformations \cite{Cordova:2018cvg}. This is very similar to what happened in the case of the chiral anomaly, and we will provide a similar interpretation in terms of symmetry fractionalization in the next section.

Let us show that the theory in \eqref{eq:SSB Abelian 2group} reproduces the 2-group symmetry \cite{Cordova:2018cvg}. First, notice that the gauge transformation
\be
\delta \Phi_0 = \lambda_0 \,, \qquad\qquad \delta \cA_1 = d\lambda_0 \,, \qquad\qquad \delta \cC_2 = d\eta_1 + \frac{k}{2\pi} \, d\lambda_0 \wedge \cA_1
\ee
leaves the action invariant. This is indeed the background gauge transformation for a 2-group. Second, in the presence of a background the currents get modified to:%
\footnote{For a $U(1)$ $p$-form symmetry we use the convention that the current $J_{p+1}$ is defined by $\star \, J_{p+1} = -i\frac{\delta S}{\delta \cA_{p+1}}$ where $\cA_{p+1}$ is the background field.}
\be
J_1 = \frac{iR^2}{2\pi} \, \bigl( d\Phi_0 - \cA_1 \bigr) + \frac{k}{4\pi} \star \bigl( d\Phi_0 \wedge da_1 \bigr) \,, \qquad\qquad J_2 = \frac{1}{2\pi} \star da_1 \,,
\ee
and these satisfy modified conservation equations
\be
d \star J_1 + \frac{k}{2\pi} \, d\cA_1 \wedge \star \, J_2 = 0 \,, \qquad\qquad d \star J_2 = 0 \,,
\ee
that are the correct conservation equations for a 2-group symmetry.

\subsection{Abelian 2-groups in the broken phase}
\label{sec:2-group SSB}

The unusual coupling to the background $\cA_1$ in \eqref{eq:SSB Abelian 2group}, responsible for the 2-group structure of the symmetry, is quite similar to the coupling \eqref{eq:symm fract coupling} responsible for a chiral anomaly, that we interpreted in terms of symmetry fractionalization. Indeed we can give a similar interpretation here too. While it is intuitively clear why symmetry fractionalization can induce a pure anomaly, and this fact has been studied extensively \cite{Delmastro:2022pfo, Brennan:2022tyl}, the necessity of symmetry fractionalization to match higher-group structures has not been much appreciated. There is indeed one important difference, namely the nature of the symmetry used to fractionalize the $U(1)$ 0-form symmetry in question: it is a \emph{composite symmetry} \cite{Brauner:2020rtz}.

In general, if we have two $U(1)$ symmetries of degrees $p$ and $q$ with currents $J_{p+1}$ and $J_{q+1}$ respectively, if $p+q \geq d-1$ we can construct a third $U(1)$ symmetry simply because the current
\be
J_{p+q-d+2} = \star \, \bigl( ( \star \, J_{p+1}) \wedge ( \star \, J_{q+1}) \bigr)
\ee
is automatically conserved. This symmetry is of degree $p+q-d+1$. In general, it is not a particularly interesting symmetry because its consequences are already implied by the constituent symmetries. However, it plays a role in our discussion. The IR theory of a 4d compact boson has an emergent 2-form symmetry: the winding symmetry of the scalar with current $J_3 = - \frac{1}{2\pi} \star d\Phi _0$. This is the symmetry we used to fractionalize the 0-form symmetry in the case of the chiral anomaly. In this case, since we also have the magnetic 1-form symmetry of the photon with current $J_2 = \frac{1}{2\pi} \star da_1$, we can construct 
\be
\widehat{J}_1 = \star \, \bigl( ( \star \, J_3) \wedge ( \star \, J_2) \bigr) = \frac{1}{4\pi^2} \star \bigl( d \Phi_0 \wedge da_1 \bigr)
\ee
that generates a 0-form symmetry. Using this symmetry to fractionalize the shift symmetry of the compact boson, as described in Section~\ref{sec:anomaly matching SSB}, we obtain precisely the non-canonical coupling in \eqref{eq:SSB Abelian 2group}.

%%%%%%%%%%%%%%%%%%%%%%%%%%%%%%%%%%%%%%%%%%%%%%%%%%%%

\section{Boundary Chern--Simons-like terms}
\label{sec:boundary chern-simons}

In this section we study bulk models obtained by adding terms without derivatives. These do not affect the boundary terms in the variational principle and hence do not modify the boundary conditions. Thus the dual theory couples minimally to the background fields, but it contains extra interactions, typically Chern--Simons-like terms. Our main motivation here is to verify our conjecture in a case with a non-invertible symmetry, the $\bQ/\bZ$ chiral symmetry in four dimensions \cite{Choi:2022jqy, Cordova:2022ieu},%
\footnote{See \cite{Arbalestrier:2024oqg} for a recent proposal to recover the full $U(1)$ chiral symmetry.}
and to provide a framework to study aspects of its spontaneous breaking. We also consider in Section~\ref{sec:maxwell chern-simons} a bulk 4d TQFT introduced in \cite{Antinucci:2024zjp}, which was argued to be related to 3d gauge theories with Chern--Simons interactions. We use our formalism to establish a precise holographic duality confirming the expectation of \cite{Antinucci:2024zjp}.

\subsection{Holographic dual to Maxwell--Chern--Simons theory}
\label{sec:maxwell chern-simons}

We consider the 4d TQFT with action
\be
S = \frac{i}{2\pi} \int_{X_4} \biggl( A_1 \wedge db_2 + \frac{\phi}{4\pi} \, b_2 \wedge b_2 \biggr) \,,
\ee
where $b_2$ is an $\bR$ 2-from gauge field, $A_1$ is a standard $U(1)$ gauge field, and $\phi$ is a parameter. On closed manifolds the theory is invariant under the following gauge transformations:
\be
\delta A_1 = d\rho_0 - \frac{\phi}{2\pi} \, \lambda _1 \,, \qquad\qquad \delta b_2 = d\lambda_1 \,.
\ee
The gauge-invariant operators include surfaces $U_\alpha(\gamma _2) = e^{i \alpha \int_{\gamma_2} b_2}$ and the generically non-genuine lines $W_n(\gamma_1, D_2) = e^{in \int_{\gamma_1} A_1 + \frac{in\phi}{2\pi} \int_{D_2} b_2}$ that need an attached two-disk $D_2$ bounded by $\gamma_1$. The label $\alpha \sim \alpha +1$ is circle valued, while $n \in \bZ$. The coupling $\phi$ is $2\pi$ periodic.

We will be mostly interested in the case
\be
\phi = \frac{2\pi}{k} \qquad\qquad \text{with} \qquad k \in \bZ \,.
\ee
In this case the lines $W_{mk}$ become genuine, and an interesting Lagrangian algebra%
\footnote{A more natural Lagrangian algebra consists of all surfaces $U_\alpha$. Used in SymTFT it describes an exotic $\bZ$ 1-form symmetry with anomaly parametrized by $\phi$ \cite{Antinucci:2024zjp}, while holographically we expect it to describe its breaking.}
is obtained by taking all the genuine lines together with the surfaces $U_{l/k}$ with $l \in \bZ_k$. Used in SymTFT \cite{Antinucci:2024zjp}, this Lagrangian algebra describes the symmetry $U(1)_{\phantom{k}}^{[0]} \times \bZ_k^{[1]}$: the first factor is a 0-form symmetry, the second factor is an anomalous 1-form symmetry (with coefficient 1), and there is a mixed anomaly between the two.

We place the theory on a manifold with boundary, where we impose the boundary condition
\be
\label{eq:bc CS}
\star \, \bigl( A_1 - \cA_1 \bigr) = - \frac{i \pi}{k^2 e^2} \, b_2 \,.
\ee
In order to have a good variational principle we must add the boundary term
\be
S_\partial = - \frac{k^2e^2}{4\pi^2} \!\int_{\partial X_4} A_1 \wedge \star \, A_1 = \frac{1}{4k^2e^2} \!\int_{\partial X_4} \biggl( b_2 + \frac{ik^2e^2}{\pi} \star \cA_1 \biggr) \wedge \star \, \biggl( b_2 + \frac{ik^2e^2}{\pi} \star \cA_1 \biggr) \,.
\ee
The gauge transformation $\delta A_1 = d\rho_0$ is restored by $\delta \cA_1 = d\rho_0$ that makes \eqref{eq:bc CS} invariant. The full system is gauge invariant, provided that we also add a counterterm $S_\text{c.t.} = \frac{k^2e^2}{4\pi^2} \!\int_{\partial X_4} \cA_1 \wedge \star \, \cA_1$.
    
We take the bulk to be the product of a three-dimensional ball $B_3$ and the time circle $S^1$, so that $\partial X_4 \equiv \cM_3 = S^2 \times S^1$. Integrating out the time components $A_0^t$, $b_1^t$ we get delta functions imposing
\be
\tilde{d} \,\widetilde{b}_2 = 0 \,, \qquad\qquad \tilde{d} \widetilde{A}_1 + \tfrac{1}{k} \, \widetilde{b}_2 = 0 \,,
\ee
that are solved introducing $\Phi_0$ and $\hat{a}_1$ through
\be
\widetilde{b}_2 = \tilde{d} \, \hat{a}_1 \,, \qquad\qquad \widetilde{A}_1 = \tilde{d} \Phi_0 - \tfrac{1}{k} \, \hat{a}_1 \,.
\ee
With this, the bulk path integral reduces to a boundary path integral with action
\begin{align}
\label{eq:luttinger maxwell chern simons}
S + S_\partial + S_\text{c.t.} &= \frac{i}{2\pi} \int_{\cM_3} \biggl[ \partial_t \hat{a}_1 \wedge \Bigl( \tilde{d} \Phi_0 - \tfrac1k \, \hat{a}_1 \Bigr) \wedge dt + \frac{1}{2k}\, \hat{a}_1 \wedge d\hat{a}_1 + \tilde{d} \, \hat{a}_1 \wedge \cA_0^t \, dt \\ 
&\quad - \frac{i}{2} \biggl( \frac{\pi}{k^2e^2} \, \tilde{d} \, \hat{a}_1 \wedge \tilde{d} \, \hat{a}_1 + \frac{k^2e^2}{\pi} \Bigl( \tilde{d} \Phi_0 - \widetilde{\cA}_1 - \tfrac{1}{k} \, \hat{a}_1 \Bigr) \wedge \star \, \Bigl( \tilde{d} \Phi_0 - \tilde{\cA}_1 - \tfrac{1}{k} \, \hat{a}_1 \Bigr) \biggr) \biggr] \,. \nn
\end{align}
Attempting to integrate out $\hat{a}_1$ to derive a covariant action for the scalar field, as we did in Section~\ref{sec: 0-form goldstone}, results in a non-local action.%
\footnote{A similar (even though less transparent) problem would have arisen if we tried to obtain the boundary theory using the second method described at the end of Section~\ref{sec: 0-form goldstone}, \ie, by integrating out directly the whole $b_2$: the latter does not appear linearly in the bulk action.}
However, there is no problem in integrating out $\Phi_0$ from \eqref{eq:luttinger maxwell chern simons} and we obtain a local and covariant boundary theory with action
\be
S = \frac{1}{4k^2e^2} \int_{\cM_3} d \hat{a}_1 \wedge \star \,  d\hat{a}_1 + \frac{i}{4\pi k} \int_{\cM_3} \hat{a}_1 \wedge d \hat{a}_1 + \frac{i}{2\pi} \int_{\cM_3} d \hat{a}_1 \wedge \cA_1 \,.
\ee

This might seem like a $U(1)$ gauge theory with an improperly quantized Chern--Simons level. However we must be careful in identifying the correct $U(1)$ gauge field, by considering the condensation of the Lagrangian algebra that trivializes the bulk. This includes all genuine lines as well as $k$ surfaces:
\be
\cL = \Bigl\{ W_{km} = e^{i k m \int\! A_1} \,,\, U_{l/k} = e^{\frac{il}{k} \int b_2} \Bigm| m \in \bZ \,,\;\; l \in \bZ_k \Bigr\} \,.
\ee
On the geometry that we are considering, condensing $\cL$ amounts to inserting the lines $W_{km}$ along the time circle and summing over $m$, while the surfaces have no effect. The insertion of $W_{km}$ modifies the path integral so as to impose that $\int_{S^2} b_2 = 2\pi k m$ for any two-sphere in $B_3$ that surrounds the Wilson line. This in particular includes the boundary spatial manifold. From the boundary theory viewpoint, this is a topological sector of the path integral with flux
\be
\int_{S^2} \frac{d \hat{a}_1}{2\pi} = km \,.
\ee
Hence the canonically normalized $U(1)$ gauge field is $a_1 = \hat{a}_1 / k$, in terms of which the boundary theory has action
\be
S = \frac{1}{4e^2} \int_{\cM_3} da_1 \wedge \star \, da_1 + \frac{ik}{4\pi} \int_{\cM_3} a _1 \wedge da_1 + \frac{ik}{2\pi} \int_{\cM_3} d a_1 \wedge \cA_1 \,.
\ee
This is Maxwell--Chern--Simons theory at level $k$, coupled to a background field for the topological $U(1)$ symmetry acting on monopoles. More precisely, the background field for this symmetry is $\cA_1' = k \cA_1$, while $\cA_1$ is the background for a larger non-faithful $U(1)$ symmetry obtained by extending the topological symmetry with a trivially-acting $\bZ_k$.%
\footnote{The reason why we got this coupling is that the TQFT we started with describes this larger symmetry, implemented by the operators $e^{i\alpha \!\int\! b_2}$, but the subgroup $\bZ_k$ was condensed in the bulk, and acts trivially in the boundary theory. As discussed in Section~\ref{sec: 0-form goldstone}, we did not explicitly introduce a background for the $\bZ_k$ 1-form symmetry.}

It should be noted that this example has a slightly different flavor than all other ones discussed in this paper. The UV symmetry is $U(1)_{\phantom{k}}^{[0]} \times \bZ_k^{[1]}$, however only $\bZ_k$ is spontaneously broken, indeed there are no Goldstone bosons in the IR since the photon is massive due to the Chern--Simons term. We consider this example as a warm up for the next one.

\subsection[Spontaneously broken non-invertible \tps{$\bQ/\bZ$}{Q/Z} chiral symmetry]{Spontaneously broken non-invertible \tps{\matht{\bQ/\bZ}}{Q/Z} chiral symmetry}
\label{sec:Q/Z}

In 4d theories of massless Dirac fermions coupled with dynamical $U(1)$ gauge fields (QED-like theories) the classically preserved axial symmetry $U(1)_A$ suffers from an ABJ anomaly that spoils the conservation of its current: $d \star J^{(A)}_1 = \frac{k}{8\pi^2} \, F_2 \wedge F_2$ \cite{Adler:1969gk, Bell:1969ts}. Traditionally, this was interpreted as the absence of $U(1)_A$ in the quantum theory. Recently \cite{Choi:2022jqy, Cordova:2022ieu} showed that axial transformations labelled by rational numbers survive at the quantum level, but they obey non-invertible fusion rules. The SymTFT for this non-invertible chiral symmetry was derived in \cite{Antinucci:2024zjp}:
\be
\label{eq:Q/Z symTFT}
S = \frac{i}{2\pi} \int_{X_5} \biggl( b_3 \wedge dA_1 + f_2 \wedge dG_2 + \frac{k}{4\pi} \, A_1 \wedge f_2 \wedge f_2 \biggr) \,.
\ee
Here $A_1$, $G_2$ are $U(1)$ gauge fields, while $b_3$, $f_2$ are $\bR$ gauge fields. The  gauge transformations are
\bea
\delta A_1 &= d\rho_0 \,, \qquad & \delta b_3 &= d \xi_2 - \frac{k}{4\pi} \, \lambda_1 \wedge d\lambda_1 - \frac{k}{2\pi} \, \lambda_1 \wedge f_2 \,, \\
\delta f_2 &= d\lambda_1 \,, \qquad & \delta G_2 &= d\eta_1 - \frac{k}{2\pi} \, \rho_0 \, \bigl( f_2 + d\lambda_1 \bigr) - \frac{k}{2\pi} \, \lambda_1\wedge A_1 \,.
\eea
As shown in \cite{Antinucci:2024zjp}, the gauge-invariant genuine topological defects are:
\bea
W_n(\gamma_1) &= e^{in \int_{\gamma_1} A_1} \,,\qquad & U_{\frac{p}{kq}}(\gamma_3) &= e^{i \frac{p}{kq} \int_{\gamma_3} b_3} \, \cA^{q,p}(\gamma_3; f_2) \,, \\
V_\alpha(\gamma_2) &= e^{i \alpha \int_{\gamma_2} f_2} \,, \qquad & \cT_m(\gamma_2) &= e^{im \int_{\gamma_2} G_2} \, \bZ_{km}(\gamma_2; A_1, f_2) \,.
\eea
Here $n,m\in \bZ$ and $\alpha \in \bR/\bZ$, while $p/q \in \bQ$ with $\text{gcd}(p,q)=1$ and $p\sim p+kq$ so that the label $p/kq  \in \bQ/\bZ$. Then $\bZ_{km}(\gamma_2; A_1, f_2)$ denotes a pure 2d $\bZ_{km}$ gauge theory on $\gamma_2$, whose 0-form and 1-form symmetries are coupled, respectively, to $A_1$ and $f_2$. Similarly, $\cA^{q,p}(\gamma_3;f_2)$ is the minimal Abelian TQFT with $\bZ_q$ 1-form symmetry and anomaly labeled by $p$ introduced in \cite{Hsin:2018vcg}, whose 1-form symmetry is coupled to $f_2$. Stacking these TQFTs is necessary in order to make the operators gauge invariant and topological. The theories $\cA^{q,p}$ are nontrivial for any $q \neq 1$, so that only a $\bZ_k$ subgroup of the operators $U_{\frac{p}{kq}}$ (those with $q=1$) are invertible, while all other ones obey non-invertible fusion rules. Similarly, $\cT_m$ are non-invertible. In the SymTFT approach it is natural to choose topological boundary conditions associated with the Lagrangian algebra
\be
\cL = \Bigl\{ W_n \,,\, \cT_m \Bigm| n,m \in \bZ \Bigr\} \,.
\ee
The remaining operators $U_{\frac{p}{kq}}(\gamma_3)$ and $V_\alpha(\gamma_2)$ implement the non-invertible symmetry and the magnetic 1-form symmetry, respectively.

Continuing with the approach we have followed so far, we want to consider a theory of gravity based on \eqref{eq:Q/Z symTFT} with the condensation of $\cL$ in the bulk. We place this theory on a manifold $X_5$ with a boundary and impose the non-topological boundary conditions
\be
\label{eq:bc Q/Z}  
\star \, A_1 = - \frac{i}{R^2} \, b_3 + \star \, \cA_1 \,, \qquad\qquad \star \, G_2 = - \frac{i\pi}{e^2} \, f_2 + \star \, \cG_2 \,.
\ee
We need to add a boundary term:
\be
\label{eq:bdry term axion}
S_\partial = - \frac{1}{4\pi R^2} \int_{\partial X_5} b_3 \wedge \star \, b_3 - \frac{1}{4e^2} \int_{\partial X_5} f_2 \wedge \star \, f_2 \,.
\ee
As before we would like to assign gauge transformation rules to the boundary fields $\cA_1$, $\cG_2$ in order to restore some of the gauge transformations on the boundary, corresponding to the symmetries that become global there. However, while we can restore $\delta G_2 = d\eta_1$ by transforming $\delta \cG_2 = d\eta_1$, the gauge transformation $\delta A_1=d\rho _0$ cannot be restored. Indeed, while the first eqn. in \eqref{eq:bc Q/Z} could be made gauge invariant by prescribing that $\delta \cA_1 = d\rho_0$, the second one would not be invariant because $G_2$ transforms as $\delta G_2 = - \frac{k}{2\pi} \rho _0 f_2$. This term cannot be reabsorbed by modifying the gauge transformations of $\cG_2$, since $f_2$ is a dynamical field. Thus the only way to make the boundary conditions gauge invariant is to freeze the boundary value of $\rho_0$, as those of $\lambda _1$ and $\xi _2$. 

To get the boundary theory, as before, we integrate out the time components imposing
\be
\tilde{d} \widetilde{A}_1 = 0 \,, \qquad \tilde{d} \widetilde{f}_2 = 0 \,, \qquad \tilde{d} \, \widetilde{b}_3 + \frac{k}{4\pi} \, \widetilde{f}_2 \wedge \widetilde{f}_2 = 0  \,, \qquad \tilde{d} \, \widetilde{G}_2 + \frac{k}{2\pi} \, \widetilde{A}_1 \wedge \widetilde{f}_2 = 0 \,,
\ee
which are solved by
\be
\widetilde{A}_1 = \tilde{d} \Phi _0 \,, \qquad \widetilde{f}_2 = \tilde{d} a_1 \,, \qquad \widetilde{b}_3 = \tilde{d} \omega_2 - \frac{k}{4\pi} \, a_1 \wedge \tilde{d} a_1 \,, \qquad \widetilde{G}_2 = \tilde{d} C_1 - \frac{k}{2\pi} \, \Phi_0 \, \tilde{d} a_1 \,.
\ee
The total action reduces to a boundary theory with action:
\begin{align}
S &= \frac{i}{2\pi} \int_{\cM_4} \biggl[ \Bigl( \tilde{d}\omega_2 - \tfrac{k}{4\pi} \, a_1 \wedge \tilde{d} a_1 \Bigr) \wedge \bigl( \partial_t \Phi_0 - \cA_0^t \bigr) dt -\Bigl( \tilde{d} C_1 - \tfrac{k}{2\pi} \, \Phi_0 \, \tilde{d} a_1 \Bigr) \wedge \partial_t a_1 \wedge  dt \nn \\
& - \frac{i}{2} \biggl( R^2 \, \Bigl( \tilde{d} \Phi_0 - \widetilde{\cA}_1 \Bigr) \wedge \star \, \Bigl( \tilde{d} \Phi_0 - \widetilde{\cA}_1 \Bigr) + \frac{1}{R^2} \Bigl( \tilde{d}\omega_2 - \tfrac{k}{4\pi} \, a_1 \wedge \tilde{d} a_1 \Bigr) \wedge \star \, \Bigl( \tilde{d} \omega_2 - \tfrac{k}{4\pi} \, a_1 \wedge \tilde{d} a_1 \Bigr) \biggr) \nn \\ 
& - \frac{i}{2} \biggl( \frac{\pi}{e^2} \, \tilde{d} a_1 \wedge \star \, \tilde{d} a_1 + \frac{e^2}{\pi} \, \Bigl( \tilde{d} C_1 - \tfrac{k}{2\pi} \, \Phi_0 \, \tilde{d} a_1 - \widetilde{\cG}_2 \Bigr) \wedge \star \, \Bigl( \tilde{d} C_1 - \tfrac{k}{2\pi} \, \Phi_0 \, \tilde{d} a_1 - \widetilde{\cG}_2 \Bigr) \biggr) \nn \\
& + \tilde{d} a_1 \wedge \cG_1^t \wedge dt + \frac{ik}{4\pi} \, \Phi_0 \, da_1 \wedge da_1 \biggr]
\end{align}
where $\cM_4 = \partial X_5$. We can then integrate out both $\omega _2$ and $C_2$ obtaining
\be
\label{eq:axion maxwell}
S = \int_{\cM_4} \biggl[ \frac{R^2}{4\pi} \, \bigl( d\Phi_0 - \cA_1 \bigr) \wedge \star \, \bigl( d\Phi_0 - \cA_1 \bigr) + \frac{1}{4e^2} \, da_1 \wedge \star \, da_1 + \frac{ik}{8\pi^2} \, \Phi_0 \, da_1 \wedge da_1 + \frac{i}{2\pi} \, da_1 \wedge \cG_2  \biggr] \,.\;\;
\ee
As in the cases of the Abelian 2-group and of Maxwell--Chern--Simons theory, gauging the Lagrangian algebra introduces fluxes for $a_1$ turning it into a standard $U(1)$ gauge field. The theory in (\ref{eq:axion maxwell}) describes a compact boson $\Phi_0$ and a photon $a_1$ interacting via an axion coupling. This is called axion-Maxwell theory, and the full structure of its symmetries (including some emergent ones) has been studied in great detail in \cite{Choi:2022fgx}. From the discovery of the non-invertible chiral symmetry, it has been suspected that axion-Maxwell theory universally describes its symmetry breaking \cite{Cordova:2022ieu, Choi:2022fgx, Pace:2023mdo}.%
\footnote{For instance, the 4d $\bC\bP^1$ non-linear sigma model enjoys a $\bQ/\bZ$ non-invertible symmetry \cite{Chen:2022cyw} and it was argued in \cite{Pace:2023mdo} that its breaking leads to axion-Maxwell theory.}
Our result confirms that. Notably, this is the first interacting boundary theory we found among the examples considered so far.

Some comments on the coupling to the background fields are in order. As we already noticed after \eqref{eq:bdry term axion}, there is no sensible gauge transformation rules that we could assign to $\cA_1$ and $\cG_2$ to make the boundary condition invariant under $\delta A_1= d\rho _0$, hence we needed to freeze it. In the action \eqref{eq:axion maxwell}, $\cA_1$ should not be thought of as the background field for the 0-form non-invertible symmetry, but rather just as an external source that couples with the operator $J^{(A)}_1$. This is enough for holography, but it might seem a bit unsatisfactory from a symmetry viewpoint. However, this is really the hallmark of the non-invertible nature of the symmetry: ordinary background gauge fields seem not to exist, and they are effectively replaced by boundary values of dynamical fields in one dimension higher \cite{Antinucci:2022vyk}. The underlying reason is that non-invertible symmetries map untwisted sectors to twisted sectors, hence the gauge transformations of a background gauge field necessarily involve an interplay among backgrounds that do not exist simultaneously in the theory, but only in the SymTFT (or in holography) where all global variants are on the same footing. This is the reason why SymTFT is the main tool for discussing anomalies \cite{Kaidi:2023maf, Zhang:2023wlu, Antinucci:2023ezl, Cordova:2023bja}.

%%%%%%%%%%%%%%%%%%%%%%%%%%%%%%%%%%%%%%%%%%%%%%%%%%%%

\section{Non-Abelian Goldstone bosons}
\label{sec:non-Abelian}

A very interesting class of examples are those of spontaneously broken non-Abelian symmetries. In these cases the boundary EFTs that we derive are interacting and generically non-renormalizable. In the 2d/3d case we will be able to recover and somewhat generalize the CS/WZW correspondence outside of the conformal point, while in higher dimensions we will obtain the pion Lagrangian on the boundary. We start with the non-Abelian generalization of the theories considered in Section~\ref{sec: U(1) goldstone} and then add an anomaly term, which corresponds to WZW terms in various dimensions. Finally we show how our setup is able to produce an EFT for spontaneously broken non-Abelian 2-group symmetries.

\subsection{Holographic dual to the pion Lagrangian}

Let $G$ be a connected and compact Lie group (with Lie algebra $\fg$). The SymTFT for a non-Abelian
0-form symmetry $G$ in $d$ dimensions is the TQFT with action \cite{Antinucci:2024zjp, Brennan:2024fgj, Bonetti:2024cjk}:%
\footnote{For $d=3$ this theory was first considered by Horowitz \cite{Horowitz:1989ng}. Curiously, the motivation was precisely to view it as an exactly solvable theory of gravity.}
\be
\label{non-Abelian BF action}
S = \frac{i}{2\pi} \int_{X_{d+1}} \Tr \bigl( b_{d-1} \wedge F_2 \bigr) \,,
\ee
where $F_2 = dA_1 + i A_1 \wedge A_1$ is the field strength of a $G$ connection $A_1$ while $b_{d-1}$ is a $\mathfrak{g}$-valued $(d-1)$-form. The gauge transformations are
\be
A_1 \,\mapsto\, \Lambda \, A_1 \, \Lambda^{-1} + i \, d\Lambda \, \Lambda^{-1} \,, \qquad\qquad b_{d-1} \,\mapsto\, \Lambda \, b_{d-1} \, \Lambda^{-1}
\ee
as well as
\be
\label{eq:gauge transf b}
b_{d-1} \,\mapsto\, b_{d-1} + D_A \lambda_{d-2}  \,.
\ee
Here $D_A = d + i [A_1, \,\cdot\; ]_{\pm}$ is the covariant derivative that acts on $p$-forms valued in the Lie algebra as
\be
D_A \eta_p = d\eta_p + i \bigl( A_1 \wedge \eta_p - (-1)^p \, \eta_p \wedge A_1 \bigr) \,.
\ee
The topological defects of this TQFT include the Wilson lines 
\be
W_\fR (\gamma_1) = \Tr_\fR \operatorname{Pexp} \biggl( i \int_{\gamma_1} A_1 \biggr)
\ee
labelled by the irreducible representations $\fR$ of $G$, as well as $(d-1)$-dimensional Gukov--Witten operators $U_{[g]}(\gamma _{d-1})$ labelled by conjugacy classes $[g]$ of $G$ and defined by prescribing that the holonomy of $A_1$ around $U_{[g]}$ be in $[g]$ \cite{Gukov:2006jk}.
The two classes of operators have a canonical linking given by the character $\chi_\fR([g])$. A natural Lagrangian algebra that we will condense consists of the Wilson lines in all representations of $G$.

We use the following non-topological  boundary condition and boundary term on $\cM_d=\partial X_{d+1}$:
\be
\label{eq:bcnonAbelian}
\star \, \bigl( A_1 - \cA_1 \bigr) = - \frac{i}{f_\pi^2} \, b_{d-1} \,, \qquad\qquad   S_\partial = - \frac{1}{4\pi f_\pi^2} \int_{\cM_d} \Tr \bigl( b_{d-1} \wedge \star \, b_{d-1} \bigr) \,.
\ee
We can recover the gauge transformations on the boundary by assigning the transformation rule $\cA_1 \mapsto \Lambda \cA_1 \Lambda ^{-1} + i d\Lambda \Lambda ^{-1}$ so that $\cA_1$ is interpreted as a background field for a global symmetry $G$.%
\footnote{Differently from the Abelian case, here we cannot turn on another background to rescue the other gauge symmetry as well. The reason is that the gauge transformation \eqref{eq:gauge transf b} of $b_{d-1}$ cannot be reabsorbed in the boundary condition by replacing $b_{d-1}$ with $b_{d-1} - \cB_{d-1}$ and assigning a transformation rule to $\cB_{d-1}$. Indeed, this transformation would necessarily involve the dynamical field $A_1$, instead of the background $\cA_1$.}
We can proceed with the usual steps to derive the dual boundary theory. Taking the spacetime to be $X_{d+1} = B_d \times S^1$, the path integral over time components imposes
\be
\widetilde{F}_2 = 0 \,, \qquad\qquad D_{\widetilde{A}_1} \widetilde{b}_{d-1} = 0 \,.
\ee
The first equation can be solved in terms of a $G$-valued scalar field $U$ as
\be
\widetilde{A}_1 = i \, \tilde{d}U \, U^{-1} \,.
\ee
To solve the second one, since the covariant derivative with respect to a flat connection squares to zero (\ie, it becomes a differential), we set
\be
\widetilde{b}_{d-1} = \widetilde{D} \omega_{d-2}
\ee
where $\omega_{d-2}$ is a $\mathfrak{g}$-valued $(d-2)$-form, and $\widetilde{D}$ denotes the covariant derivative with respect to $i \, \tilde{d}U \, U^{-1}$. By plugging these back, the theory reduces to a boundary action:
\begin{align}
\label{eq:LutingerliquidnonAbelian}
S &= (-1)^{d} \frac{i}{2 \pi} \int_{\cM_d} \Tr \biggl[ \widetilde{D} \omega_{d-2} \wedge \Bigl( i \, \partial_t U \, U^{-1} - \cA_0^t \Bigr) dt \biggr] \\
&\quad + \frac{1}{4\pi} \int_{\cM_d} \Tr \biggl[ \frac{1}{f_{\pi}^2} \, \widetilde{D} \omega_{d-2} \wedge \star \, \widetilde{D} \omega_{d-2} + f_{\pi}^2 \Bigl( i \, \tilde{d}U \, U^{-1} - \widetilde{\cA}_1 \Bigr) \wedge \star \, \Bigl( i \, \tilde{d}U \, U^{-1} - \widetilde{\cA}_1 \Bigr) \biggr] \,. \nn
\end{align}
One important difference with respect to the Abelian case is that $U$ and $\omega_{d-2}$ do not appear symmetrically. While $U$ appears in a complicated way, the action is still quadratic in $\omega_{d-2}$ that can thus be integrated out using its equation of motion
\be
\widetilde{D} \Bigl( \partial_t U \, U^{-1} + i \cA_0^t \Bigr) \wedge dt + \frac{(-1)^{d-1}}{f_\pi^2} \, \widetilde{D} \star \widetilde{D} \omega_{d-2} = 0 \,.
\ee
Eliminating a zero-mode as in the Abelian case, we can use this equation to determine $\widetilde{D} \omega_{d-2}$, and we find the manifestly covariant form of the boundary theory:
\be
\label{eq:pion lagrangian}
S = \frac{f_\pi^2}{4\pi} \int_{\cM_d} \Tr \biggl[ \Bigl( i \, dU \, U^{-1} - \cA_1 \Bigr) \wedge \star \, \Bigl( i \, dU \, U^{-1} - \cA_1 \Bigr) \biggr] \,.
\ee
This describes a sigma model with target $G$, coupled to a background field $\cA_1$ for the symmetry $G$ that acts as $U \mapsto g U$ with $g \in G$.
The sigma model is a non-renormalizable theory that provides the leading universal term in an expansion in number of derivatives (in 4d this is chiral perturbation theory), describing the EFT of any theory with spontaneously broken symmetry $G$ \cite{Coleman:1969sm, Callan:1969sn}.

\subsection{Non-Abelian chiral anomaly}

For any even $d$ we can add a Chern--Simons term to the bulk theory (\ref{non-Abelian BF action}):%
\footnote{Here we assume $G$ to be simple and simply connected.} 
\be
S_\text{CS} = \frac{i \kappa_d}{2\pi} \int_{X_{d+1}} \!\! \Tr \bigl( \text{CS}_{d+1}(A_1) \bigr) \,, \qquad\qquad \kappa_d = \frac{k}{(2\pi)^{\frac{d}{2}-1} \bigl( \frac{d}{2} + 1 \bigr)! } \,, \qquad\qquad k \in \bZ \,,
\ee
that describes the presence of a perturbative anomaly for $G$. In this case, differently from the Abelian one, anomaly matching requires a WZW term in the spontaneously broken phase \cite{Witten:1983tw}. We want to show that this fact is implied by our conjecture. We also consider the case of $d=2$ where, strictly speaking, our conjecture does not apply because there is no spontaneous breaking of a continuous symmetry in two dimensions.

\subsubsection*{Two dimensions}

In the case of $d=2$, we use the boundary condition 
\be
\star \, \bigl( A_1 - \cA_1 \bigr) = - \frac{i}{f_\pi^2} \, \biggl( b_1 + \frac{k}{2} \bigl( A_1 - \cA_1 \bigr) \biggr)
\ee
that is gauge invariant under $A_1 \mapsto \Lambda A_1 \Lambda ^{-1} + id\Lambda \, \Lambda ^{-1}$, $\cA_1 \mapsto \Lambda \cA_1 \Lambda ^{-1} + id\Lambda \, \Lambda ^{-1}$, and add the boundary term
\be
S_\partial = - \frac{1}{4\pi f_{\pi}^2} \int_{\partial X_3} \Tr \biggl[ \Bigl( b_1 + \tfrac{k}{2} \, A_1 \Bigr) \wedge \star \, \Bigl( b_1 + \tfrac{k}{2} \, A_1\Bigr) \biggr]
\end{equation}
to make the variational principle well defined. 

As a preliminary consistency check, we compute the gauge variation. The total gauge-transformed action differs by
\be
\Delta \bigl( S + S_\partial + S_\text{c.t.} \bigr) = \frac{ik}{4\pi} \int_{\partial X_3} \Tr \bigl( \cA_1 \wedge i \Lambda ^{-1} d\Lambda \bigr) + \frac{k}{24\pi}\int_{X_3} \Tr \bigl( (i\Lambda^{-1} d\Lambda)^3 \bigr) 
\ee
from the original one.%
\footnote{Here $\ds S_\text{c.t.} = \frac{k^2}{8\pi f_\pi^2} \int_{\partial X_3} \!\! \Tr \bigl( \cA_1 \wedge \star \, \cA_1 \bigr)$ is a counterterm we add to simplify the final result.}
Upon expanding $\Lambda = \unit + \lambda_0$ and retaining only the linear order in $\lambda_0$, this reduces to the usual form of the consistent anomaly:
\be
\delta \bigl( S + S_\partial + S_\text{c.t.} \bigr) = \frac{ik}{4\pi} \int_{\partial X_3} \Tr \bigl( \cA_1 \wedge i d\lambda_0 \bigr) \,.
\ee

One can proceed in determining the dual boundary theory similarly to the non-anomalous case. Since the boundary condition is essentially the same (simply written in a different parametrization), the only difference is the bulk Chern--Simons term which gives rise to a WZW term in the boundary theory:
\begin{align}
S &= \frac{f_{\pi}^2}{4\pi} \int_{\cM_2} \Tr \Bigl[ \bigl( i \, dU \, U^{-1} - \cA_1 \bigr) \wedge \star \, \bigl( i \, dU \, U^{-1} - \cA_1 \bigr) \Bigr] + \frac{k}{12\pi} \int_{X_3} \Tr\Bigl[ \bigl( i U^{-1} dU \bigr)^3 \Bigr] \nn \\
&\quad - \frac{ik}{4\pi} \int_{\cM_2} \Tr\Bigl[ \cA_1 \wedge i \, dU \, U^{-1} \Bigr] \,.
\end{align}
We notice that there is also a non-standard coupling to the background field, that in our approach arises because of the boundary conditions, similarly to the Abelian case. Differently from that case, however, in a purely field theoretic analysis this is not interpreted as a coupling to a diagonal symmetry (since a winding symmetry is absent here), but rather it arises from the standard trial-and-error procedure to couple the $G$ symmetry to a background in the presence of the WZW term, similarly to the 4d analysis in \cite{Witten:1983tw}.

For generic values of $f_{\pi}^2$ the theory is not conformally invariant at the quantum level. However choosing $f_\pi^2 = \frac{k}{2}$ the theory has a conserved holomorphic current which generates a Kac--Moody symmetry algebra, and it displays conformal invariance \cite{Witten:1983ar}. In this case we recover a form of the CS/WZW correspondence, which is more general on one side, being valid even outside of the conformal point, but less general on the other side, since in the conformal case it automatically produces the full physical WZW model instead of its chiral halves.

\subsubsection*{Four dimensions}

In the case of $d=4$, the 5d Chern--Simons term is
\be
\label{5d CS term}
\Tr \bigl( \text{CS}_5(A_1) \bigr) = \Tr \biggl( A_1 \wedge (dA_1)^2 + \frac{3i}{2} \, A_1^3 \wedge dA_1 - \frac{3}{5} A_1^5 \biggr) \,.
\ee
As one might suspect already from the Abelian case, in order to obtain a gauge-invariant boundary condition with a consistent variational principle we need to introduce extra terms in the boundary condition that mix background and dynamical fields. We use the same iterative procedure discussed in Appendix~\ref{app:anomalousbc} for the Abelian anomaly, even though the computations are clearly more tedious here. We find the following solution. The boundary condition is
\be
\star \, \bigl( A_1 - \cA_1 \bigr) - \frac{i \kappa_4}{f_{\pi}^2} \, \biggl( \frac{1}{2} \, \bigl( \cA_1 \, \cF_2 + \cF_2 \, \cA_1 \bigr) - \frac{i}{2} \, \cA_1^3 \biggr) = - \frac{i}{f_{\pi}^2} \, \Omega_3
\ee
where $\cF_2$ is the field strength of $\cA_1$ while
\be
\Omega_3 = b_3 + \kappa_4 \biggl( F_2 \bigl( A_1 - \cA_1 \bigr) + \bigl( A_1 - \cA_1 \bigr) F_2 - \frac{i}{2} \Bigl( \bigl( A_1 - \cA_1 \bigr)^3 + \cA_1 \Bigr)^{\!3} + \frac{1}{2} \bigl( A_1 \cF_2 + \cF_2 A_1 \bigr) \biggr)
\ee
and the boundary term is
\bea
S_\partial &= - \frac{1}{4\pi f_{\pi}^2} \int_{\partial X_5} \Tr \bigl( \Omega_3 \wedge \star \, \Omega_3 \bigr) \,+\, S_\text{top} \,+\, S_\text{c.t.} \,, \\ 
S_\text{top} &= \frac{i \kappa_4}{2\pi} \int_{\partial X_5} \Tr \biggl[ \frac{1}{2} \, F_2 \, \cA_1 \, A_1 + \frac{1}{2} \, \cA_1 \, F_2 \, A_1 - \frac{i}{4} \, A_1 \, \cA_1 \, A_1 \, \cA_1 + \frac{i}{2} \, A_1^3 \, \cA_1 \biggr] .
\eea
The counterterm $S_\text{c.t.}$ is used to simplify the final expression, and it is convenient to choose it as
\be
S_\text{c.t.} = \frac{\kappa_4^2}{4\pi f_\pi^2} \int_{\partial X_5} \!\! \Tr \Bigl[ \phi(\cA_1) \wedge \star \, \phi(\cA_1) \Bigr] \,, \qquad \phi( \cA_1) = \frac{1}{2} \Bigl( \cA_1 \wedge d\cA_1 + d\cA_1 \wedge \cA_1 + i \cA_1^3 \Bigr) \,.
\ee

The boundary condition is gauge invariant under the transformation $A_1 \mapsto \Lambda A_1 \Lambda ^{-1}+id\Lambda \, \Lambda^{-1}$, $\cA_1 \mapsto \Lambda \cA_1 \Lambda^{-1} + i d\Lambda \, \Lambda^{-1}$ and one can compute the total gauge variation
\begin{align}
\Delta \bigl( S + S_\partial \bigr) &= - \frac{i\kappa_4}{2\pi} \int_{\partial X_5} \!\! \Tr \biggl[ \bigl( i \Lambda^{-1} d\Lambda \bigr) \wedge \phi(\cA_1) + \frac{i}{4} \bigl( \cA_1 \wedge i \Lambda^{-1} d\Lambda \bigr)^2  - \frac{i}{2} \bigl( i \Lambda^{-1} d\Lambda \bigr)^3 \wedge \cA_1 \biggr] \nn \\
&\quad - \frac{i\kappa_4}{20\pi} \int_{X_5} \Tr \Bigl[ \bigl( i \Lambda^{-1} d\Lambda \bigr)^5 \Bigr] \,.
\end{align}
Expanding $\Lambda = \unit + \lambda_0$ to linear order, we recover the usual form of the consistent anomaly in four dimensions:
\be
\delta \bigl( S + S_\partial \bigr) = - \frac{ik}{48\pi ^2} \int_{\partial X_5} \! \Tr \Bigl[ i d\lambda_0 \wedge \Bigl( \cA_1 \wedge d\cA_1 + d\cA_1 \wedge \cA_1 + i \cA_1^3 \Bigr) \Bigr] \,.
\ee
We can then proceed, as before, with the reduction of the action on the boundary. We find
\begin{align}
S &= \frac{f_{\pi}^2}{4\pi} \int_{\cM_4} \Tr \biggl[ \Bigl( i dU \, U^{-1} - \cA_1 \Bigr) \wedge \star \, \Bigl( i dU \, U^{-1} - \cA_1 \Bigr) \biggr] - \frac{ik}{240\pi^2} \int_{X_5} \Tr \biggl[ \bigl( i U^{-1} dU \bigr)^5 \biggr] \nn \\
&\quad +\frac{i k}{48\pi^2} \int_{\cM_4} \Tr \biggl[ i dU \, U^{-1} \wedge \Bigl( \cA_1 \wedge \cF_2 + \cF_2 \wedge \cA_1 - \cA_1^3 \Bigr) \biggr] \\
&\quad + \frac{k}{48\pi^2} \int_{\cM_4} \Tr \biggl[ \frac{1}{2} \, i dU \, U^{-1} \wedge \cA_1 \wedge i dU \, U^{-1} \wedge \cA_1 - \bigl( i dU \, U^{-1} \bigr)^3 \wedge \cA_1 \biggr] \,. \nn
\end{align}
Turning off the background gauge field $\cA_1$ we recognize a non-linear sigma model with target space $G$ with a properly normalized WZW term, that describes the dynamics of Goldstone bosons. The coupling to the background $\cA_1$ is completely fixed by the requirement of a gauge-invariant boundary condition, and correctly captures the anomaly of the non-linearly realized $G$ symmetry.

\subsection{Non-Abelian 2-group symmetries}

In 4d one can have 2-group symmetries whose 0-form part is a non-Abelian group $G$, while the 1\nobreakdash-form part is $U(1)$. These symmetry structures arise, \eg,  if one starts from a theory with a 0-form symmetry group $U(1) \times G$ with an 't~Hooft anomaly that is linear in $U(1)$ and quadratic in $G$:
\be
\label{eq:inflow 2group}
S_\text{inflow} = \frac{ik}{8\pi^2} \int_{X_5} dV_1 \wedge \Tr \biggl( A_1 \wedge dA_1 + \frac{2i}{3} \, A_1^3 \biggr) \,,
\end{equation}
and then gauges the $U(1)$ symmetry \cite{Cordova:2018cvg}. The 1-form symmetry involved in the 2-group is the magnetic symmetry of the gauged $U(1)$. The SymTFT for this non-Abelian 2-group symmetry can be  derived using the dynamical gauging procedure described  in \cite{Antinucci:2024zjp}. Indeed one starts from the SymTFT for the $U(1) \times G$ 0-form symmetry:
\be
S' = \frac{i}{2\pi} \int_{X_5} \biggl[ g_3 \wedge dV_1 + \Tr \bigl( b_3 \wedge F_2 \bigr) + \frac{k}{4\pi} \, dV_1 \wedge \Tr \Bigl( A_1 \wedge dA_1 + \frac{2i}{3} A_1^3 \Bigr) \biggr]
\ee
where $g_3$ and $V_1$ are an $\bR$ and a $U(1)$ gauge field, respectively, $b_3$ is $\mathfrak{g}$-valued and $A_1$ is a $G$ connection ($F_2$ is its field strength). Then one applies the map introduced in \cite{Antinucci:2024zjp} that implements the dynamical gauging of $U(1)$ on the boundary from the viewpoint of the SymTFT. The net effect is the replacement $dV_1 \mapsto h_2$, $g_3 \mapsto dC_2$, thus the resulting SymTFT has action
\be
\label{SymTFT for non-Abelian 2-group}
S = \frac{i}{2\pi} \int_{X_5} \biggl[ h_2 \wedge dC_2 + \Tr \bigl( b_3 \wedge F_2 \bigr) +  \frac{k}{4\pi} \, h_2 \wedge \Tr \Bigl( A_1 \wedge dA_1 + \frac{2i}{3}A_1^3 \Bigr) \biggr] \,.
\ee
The gauge transformations are:%
\footnote{Recall that the variation of the three-dimensional Chern--Simons term is:
\be
\Tr \bigl( \text{CS}_3 (A_1) \bigr) \,\mapsto\, \Tr \bigl( \text{CS}_3(A_1) \bigr) + d \Tr \bigl( A_1 \wedge i \Lambda^{-1} d\Lambda \bigr) - \frac{i}{3} \Tr\Bigl( \bigl( i \Lambda^{-1} d\Lambda \bigr)^3 \Bigr) \,.
\ee}
\bea
h_2 &\mapsto h_2 + d \xi_1 \,, \qquad & A_1 &\mapsto \Lambda A_1 \Lambda ^{-1} + i d\Lambda \, \Lambda^{-1} \,, \\
b_3 &\mapsto b_3 - \frac{k}{4\pi} \, \xi_1 \wedge F_2 \,,\qquad & C_2 &\mapsto C_2 + d \eta_1 - \frac{k}{4\pi} \Tr \bigl( A_1 \wedge i \Lambda^{-1} d\Lambda \bigr) + \frac{ik}{6\pi} \Tr \Theta_2 \,,
\eea
where $\Theta_2$ is a locally defined real 2-form with the property that $\Tr \bigl( (i \Lambda^{-1} d\Lambda )^3 \bigr) = d \Tr \Theta_2$.

Again, we can use an iterative procedure to determine a set of gauge-invariant boundary conditions together with a boundary term that provide a good variation principle. The boundary conditions are
\be
\star \, \bigl( A_1 - \cA_1 \bigr) = - \frac{i}{R^2} \biggl( b_3 + \frac{k}{4\pi} \bigl( A_1 - \cA_1 \bigr) \biggr) \,,\quad \star \, h_2 = \frac{i e^2}{\pi} \biggl( C_2 - \cC_2 - \frac{k}{4\pi} \Tr \bigl( \cA_1 \wedge A_1 \bigr) \biggr)
\ee
while the boundary term is
\bea
S_\partial &= - \frac{i}{2\pi} \int_{\partial X_5} h_2 \wedge \biggl( C_2 - \frac{k}{4\pi} \Tr \bigl( \cA_1 \wedge A_1 \bigr) \biggr) \\
&\quad - \frac{e^2}{4\pi^2} \int_{\partial X_5} \biggl( C_2 - \frac{k}{4\pi} \Tr\bigl( \cA_1 \wedge A_1 \bigr) \biggr) \wedge \star \, \biggl( C_2 - \frac{k}{4\pi} \Tr \bigl( \cA_1 \wedge A_1 \bigr) \biggr) \\
&\quad - \frac{1}{4\pi R^2} \int_{\partial X_5} \Tr \biggl[ \biggl( b_3 + \frac{k}{4\pi} \bigl( A_1 - \cA_1 \bigr) \biggr) \wedge \star \, \biggl( b_3 + \frac{k}{4\pi} \bigl( A_1 - \cA_1 \bigr) \biggr) \biggr] .
\eea
The boundary condition becomes gauge invariant by assigning the following transformations to the backgrounds $\cA_1$ and $\cC_2$:
\be
\cA_1 \mapsto \Lambda \cA_1 \Lambda^{-1} + i d\Lambda \, \Lambda^{-1} \,, \qquad \cC_2 \mapsto \cC_2 + d\eta_1 - \frac{ik}{4\pi} \Tr \bigl( \cA_1 \wedge \Lambda^{-1} d\Lambda \bigr) + \frac{ik}{12\pi} \Tr \Theta_2 \,.
\ee
These reproduce the background gauge transformation of \cite{Cordova:2018cvg} for a non-Abelian 2-group symmetry upon expanding $U = \unit + \lambda_0$ at first order:
\be
\delta \cA_1 = i D_{\cA_1}\lambda_0 \,, \qquad\qquad \delta \cC_2 =  d\eta_1 - \frac{ik}{4\pi} \Tr \bigl( \cA_1 \wedge d\lambda_0 \bigr) \,.
\ee
It is also easy to see that the whole bulk-boundary system is gauge invariant under transformations of $A_1$ and $C_2$ provided we add a counterterm $S_\text{c.t.} = \frac{e^2}{4\pi^2} \int_{\partial X_5} \cC_2 \wedge \star \, \cC_2$.

We can apply our usual machinery to get the dual boundary theory. We obtain a $G$-valued scalar field $U$ from $A_1$, and a Maxwell field $a_1$ from $h_2$, with the following boundary action:
\bea
\label{eq:EFT non-abelian 2group}
S &= \frac{f_{\pi}^2}{4\pi} \int_{\cM_4} \! \Tr \biggl[ \Bigl( i dU \, U^{-1} - \cA_1 \Bigr) \wedge \star \, \Bigl( i dU \, U^{-1} - \cA_1 \Bigr) \biggr] + \frac{1}{4 e^2} \int_{\cM_4} da_1 \wedge \star \, da_1 \\
&\quad + \frac{k}{24\pi^2} \int_{\cM_4} a_1 \wedge \Tr\Bigl[ \bigl( i U^{-1} dU \bigr)^3 \Bigr] \\
&\quad + \frac{i}{2\pi} \int_{\cM_4} da_1 \wedge \Tr\Bigl[ \cA_1 \wedge i U^{-1} dU \Bigr] + \frac{i}{2\pi} \int_{\cM_4} \cC_2 \wedge da_1 \,.
\eea
In the first line we recognize a non-linear sigma model with target space $G$ and a Maxwell theory. The last line describes the coupling to the background field $\cC_2$ for the magnetic $U(1)$ 1-form symmetry, as well as a nonstandard coupling to the background $\cA_1$ for the symmetry $G$, similar to the one arising in the Abelian case in Section~\ref{sec: 2-groups}. The most interesting new thing here is the term in the second line that describes a coupling between the photon and the pions. This is a linear coupling of the photon to the current of a topological symmetry that exists in any sigma model with target $G$. According to our conjecture, this model is the universal EFT that describes the IR of any theory with a spontaneously broken non-Abelian 2-group symmetry. To the best of our knowledge, this universal EFT was not derived elsewhere.

Some comments on the extra Wess--Zumino-like coupling are in order. First, in any RG flow that breaks the 2-group spontaneously, this coupling must be generated as a consequence of the 2-group matching. In a sense, it is similar to the presence of the WZW term in the EFT of a spontaneously broken anomalous non-Abelian symmetry. Quite like that term, it breaks a symmetry of the EFT that would be there if $k=0$. Indeed, for $k=0$ the theory is separately invariant under four $\bZ_2$ symmetries: parity $P_0\,{:}\; x_i \mapsto -x_i$ for $i=1,2,3$; photon charge conjugation $C_1\,{:}\; a_1 \mapsto -a_1$; non-Abelian charge conjugation%
\footnote{The reason for this name will be clear in the upcoming discussion of $U(N)$ QCD.}
$C_2\,{:}\; U \mapsto U^\sT$; pion number mod-2 $(-1)^{N_\pi}\,{:}\; U \mapsto U^{-1}$. All these four symmetries are violated by the photon-pion coupling, but the product of any two of them is preserved. Therefore the discrete symmetry for $k \neq 0$ is $(\bZ_2)^3 $ generated by
\be
\label{eq:symm 2-group}
P = P_0 \, (-1)^{N_\pi} \,, \qquad\qquad C = C_1 \, C_2 \,, \qquad\qquad \wt{C} = C_1 \, (-1)^{N_\pi} \,.
\ee
The photon-pion coupling allows, for instance, a process involving three pions and one photon, which would have been forbidden otherwise. We summarize the various symmetry actions and charges in Table~\ref{tab: sym 2-group}.

\begin{table}[t]
$$
\begin{array}{|c||c||c|c|c|}
\hline \rule[-1em]{0pt}{2.8em}
  & \text{Definition} & \ \ \ \   x_i \ \ \ \  &   a_1  \Tr \Bigl[ \bigl( i U^{-1} dU \bigr)^3 \Bigr] & \Tr \Bigl[ \bigl( i U^{-1} dU \bigr)^5 \Bigr] \cr
\hline \hline \rule[-1em]{0pt}{2.8em}
   P_0  & x_i\mapsto -x_i & -1 & -1 & -1 \cr
\hline \rule[-1em]{0pt}{2.8em}
    C_1 & a_1\mapsto -a_1 & 1 & -1   & 1 \cr
\hline \rule[-1em]{0pt}{2.8em}
   C_2 & U\mapsto U^\sT & 1 &   -1  & 1 \cr \hline \rule[-1em]{0pt}{2.8em}
   (-1)^{N_\pi} & U\mapsto U^{-1} & 1 &   -1 & -1 \cr \hline
\end{array}
$$
\caption{\label{tab: sym 2-group}%
The four $\bZ_2$ symmetries, and the corresponding phases acquired by the coordinates, the photon-pion coupling term, and the standard WZW term, respectively. Notice that while $\Tr \bigl[ (i U^{-1} dU)^5 \bigr]$ is invariant under $U \mapsto U^\sT$, the term $\Tr \bigl[ (iU^{-1} dU)^3 \bigr]$ changes sign.}
\end{table}

Second, the 2-group symmetry we started with could suffer from a perturbative cubic chiral anomaly for $G$ as well. This would be described by the addition of a 5d Chern--Simons term (\ref{5d CS term}) to the bulk action in (\ref{SymTFT for non-Abelian 2-group}), and would result in an extra WZW term $S_\text{WZW} = - \frac{ik}{240\pi^2} \!\int_{X_5} \! \Tr\bigl[ ( i U^{-1} dU )^5 \bigr]$ in the 4d boundary action (\ref{eq:EFT non-abelian 2group}).%
\footnote{We did not work out the detailed form of the coupling to the background fields.}
This term would further break the discrete symmetry of the EFT to $(\bZ_2)^2$ generated by $P$ and $C$, as it is clear from Table~\ref{tab: sym 2-group}.

\paragraph{An application: \tps{\matht{U(N)}}{U(N)} QCD.}
Let us present a concrete application of the effective action \eqref{eq:EFT non-abelian 2group}. Consider a 4d gauge theory with $U(N)$ gauge group and $N_f$ flavors of massless Dirac fermions, so that there is a chiral symmetry $SU(N_f)_L\times SU(N_f)_R$. It can be obtained by gauging the baryon number symmetry $U(1)_B$ in ordinary $SU(N)$ QCD, hence it contains an Abelian gauge field $A_\mu$ on top of the non-Abelian gauge fields. Being weakly coupled at low energy, $A_\mu$ is not expected to drastically modify the strong coupling dynamics of the non-Abelian sector. Hence for $N_f$ small enough, the quark bilinear takes VEV and spontaneously breaks the chiral symmetry:%
\footnote{Notice that the usual argument \cite{tHooft:1979rat} based on 't~Hooft anomaly matching in $SU(N)$ QCD is also valid here, hence we do not really need to make the assumption that the photon does not affect chiral symmetry breaking.}
\be
SU(N_f) _L\times SU(N_f) _R \;\rightarrow\; SU(N_f)_V  
\ee
producing at low energy massless pions that interact as a non-linear sigma model with target space $SU(N_f)$. The pions are neutral under the non-Abelian gauge symmetry $SU(N)$, whose gluons are confined. However the Abelian gauge field $A_\mu$ remains even in the deep IR and there is no reason why it should be decoupled from the non-linear sigma model. Indeed, while the pion fields themselves are neutral under $U(1)$, being bound states of quarks it is a priori unclear whether there is a low-energy remnant of the quark-photon interaction.

We can answer this question using our result, and showing that the photon is not decoupled. Indeed there is a $U(1)$ magnetic 1-form symmetry from the Abelian gauge field (that is its Goldstone boson), which forms a non-trivial 2-group with $SU(N_f)_L$ (and also with $SU(N_f)_R$, but we can just focus on one of the two). To see this, we notice that there is a triangle anomaly $U(1)\,$-$\,SU(N_f)_L^2$ whose anomaly polynomial is 
\be
\cP_{U(1) \,\text{-}\, SU(N_f)_L^2} = \frac{N}{8\pi^2} \, dA \wedge \Tr \bigl( \cF \wedge \cF \bigr) \,,
\ee
where $\cF = d \cG + i \, \cG \wedge \cG$ is the field strength of the background field $\cG$ for $SU(N_f)_L$. The coefficient $N$ comes because all left-moving fermions have charge $1$ under $U(1)$ and are in the fundamental representation of the non-Abelian gauge symmetry $SU(N)$. By comparison with \eqref{eq:inflow 2group} we read off that the $U(1)$ 1-form symmetry and $SU(N_f)_L$ form a 2-group with $k=N$. Because of chiral symmetry breaking and spontaneous breaking of the 1-form symmetry, the 2-groups is fully broken and, from our result above, the low-energy EFT describing pions and photon is \eqref{eq:EFT non-abelian 2group}, plus the standard WZW term (also with coefficient $N$) for the pions due to the cubic $SU(N_f)_L$ anomaly:%
\footnote{2-group structures in sigma models arising in the IR of QCD-like theories have been recently considered also in \cite{Davighi:2024zjp}. The IR there, however, is purely scalar, and the 2-group is not fully spontaneously broken (the 1-form symmetry is preserved). The interaction responsible for the 2-group is not a photon-pion coupling, but rather a coupling between pions parametrizing two different target spaces. Indeed the UV model studied in \cite{Davighi:2024zjp} can be obtained from $U(N)$ QCD by adding scalars charged under $U(1)_B$ that Higgs the Abelian gauge field.}
\bea
\label{eq:EFT U(N) QCD}
S_\text{IR} &= \frac{f_{\pi}^2}{4\pi} \int_{\cM_4} \Tr \Bigl[ \bigl( i dU \, U^{-1} \bigr) \wedge \star \, \bigl( i dU \, U^{-1} \bigr) \Bigr] + \frac{1}{4 e^2} \int_{\cM_4} dA \wedge \star \, dA \\ 
&\quad + \frac{N}{24\pi^2} \int_{\cM_4} A \wedge \Tr \Bigl[ \bigl( i U^{-1} dU \bigr)^3 \Bigr] - \frac{iN}{240\pi^2} \int_{X_5} \Tr \Bigl[ \bigl( i U^{-1} dU \bigr)^5 \Bigr] \,.      
\eea
Thus, while the pions themselves are uncharged under the $U(1)$ gauge group, the photon $A$ is coupled with an effective current
\be
J_B = - \frac{N}{24\pi^2} \star \Tr \Bigl[ \bigl( i U^{-1} dU \bigr)^3 \Bigr] .
\ee
This current is conserved, and in the absence of the pion-photon interaction it generate a global $U(1)$ symmetry of the sigma model: the topological symmetry due to the non-trivial homotopy group $\pi_3 \bigl( SU(N_f) \bigr) = \bZ$. The integral of $\star \, J_B$ gives indeed the winding number:
\be
\label{eq:winding skyrmions}
w(\cM_3) = - \frac{i}{24\pi^2} \int_{\cM_3} \Tr \Bigl[ \bigl( i U^{-1} dU \bigr)^3 \Bigr] \;\in\, \bZ \,.
\ee
In the $U(N)$ theory, configurations with nontrivial winding have a $U(1)$ gauge charge. These configurations are Skyrmions: solitonic objects which, in the $SU(N)$ theory, are identified with the baryons \cite{Witten:1983tw, Witten:1983tx}. This is confirmed by our finding: the $U(N)$ theory is obtained from ordinary $SU(N)$ QCD by gauging the baryon number symmetry, hence the baryons are no longer gauge invariant, but rather are coupled with $A$.

We can make this more precise as follows. In the absence of the photon-pion coupling, the operators charged under the topological $U(1)$ symmetry are local operators $\cB_q(x)$ defined as disorder operators which impose that 
\be
w(S^3) = q \in \bZ
\ee
on a 3-sphere $S^3$ that links with $x$. Similarly to the monopole operator in Chern--Simons theory, $\cB_q(x)$ gets a gauge charge $Nq$ due to the coupling with the photon.

Also, in the absence of the 2-group structure, the low-energy effective theory would have an emergent electric $U(1)$ 1-form symmetry shifting $A \rightarrow A+\lambda$ (with the periods of $\lambda $ in the interval $[0,2\pi]$) and acting on the Wilson lines $W_n(\gamma) = e^{i n \int _\gamma A}$. Because of the photon-pion coupling, however, only a $\bZ_N\subset U(1)$ subgroup of this 1-form symmetry emerges. Indeed using the quantization \eqref{eq:winding skyrmions}, shifting $A \rightarrow A+\lambda $ leaves the exponentiated action invariant only if the periods of $\lambda $ are multiples of $\frac{2\pi}{N}$. An equivalent way to see this is that the Wilson line $W_{n=N}$ can terminate on the Baryon operator $\cB_1(x)$. Notice that the microscopic theory does not have this $\bZ_N$ 1-form symmetry, because the quarks have unit charge under the gauged $U(1)_B$. The emergence of $\bZ_N$ has a clear interpretation: the quarks are confined and the only dynamical particles charged under $U(1)_B$ at low energy are baryons, with charges multiple of $N$.

As a final comment, notice that among the three $\bZ_2$ symmetries $P$, $C$, $\wt{C}$ defined in \eqref{eq:symm 2-group} that are preserved by the photon-pion coupling, only $P$ and $C$ are preserved also by the standard WZW term, while $\wt{C}$ is explicitly broken (see Table~\ref{tab: sym 2-group}). This has to do with the fact that in $U(N)$ QCD, $C_2\,{:}\; U \mapsto U^\sT$ is the low-energy remnant of the non-Abelian charge conjugation that, in the UV, also acts on the $SU(N)$ gauge bosons, confined in the IR. In the $U(N)$ theory this charge conjugation is not independent from the Abelian charge conjugation $C_1$ acting on the photon, since the fermions are in the fundamental representation of both. Hence, only the product $C = C_1C_2$ is a symmetry of the theory.

%%%%%%%%%%%%%%%%%%%%%%%%%%%%%%%%%%%%%%%%%%%%%%%%%%%%

\subsection*{Acknowledgments}
We are grateful to Riccardo Argurio, Matthew Bullimore, Christian Copetti, Lorenzo Di Pietro, Giovanni Galati, I\~naki Garc\'ia Etxebarria, Matteo Sacchi, Antonio Santaniello, Sakura Sch\"afer-Nameki, Marco Serone, and Matthew Yu for discussions. We are supported
by the ERC-COG grant NP-QFT No. 864583 ``Non-perturbative dynamics of quantum fields: from new deconfined phases of matter to quantum black holes",
by the MUR-FARE2020 grant No. R20E8NR3HX ``The Emergence of Quantum Gravity from Strong Coupling Dynamics",
by the MUR-PRIN2022 grant No. 2022NY2MXY,
and by the INFN ``Iniziativa Specifica ST\&FI".

\appendix

%%%%%%%%%%%%%%%%%%%%%%%%%%%%%%%%%%%%%%%%%%%%%%%%%%%%

\section{Anomalous boundary conditions}
\label{app:anomalousbc}

In this appendix we present an iterative procedure to consistently turn on a background for boundary theories with a $U(1)$ anomalous symmetry in generic even dimension. For the sake of concreteness we present this procedure in the simplest case of a $U(1)$ symmetry with anomaly, but the same idea can be used for higher groups and in the non-Abelian cases discussed in the main text. In general, the method presented here is necessary to determine consistent boundary conditions whenever the simple BF theory is modified by some non-Gaussian term containing derivatives.

Consider the TQFT with action
\be
S = \frac{i}{2\pi} \int_{X_{d+1}} \Bigl( b_{d-1} \wedge dA_1 + \kappa_d \, A_1 \wedge (dA_1)^{\frac{d}{2}} \Bigr) \,, \qquad\qquad \kappa_d = \frac{k}{ (2\pi)^{ \frac{d}{2}-1} \bigl(\frac{d}{2}+1 \bigr)!} \,,
\ee
and $k \in \bZ$. In the presence of a boundary, the variation of the action produces a term
\be
- \frac{i}{2\pi} \int_{\partial X_{d+1}} \biggl( b_{d-1} + \frac{d}{2} \, \kappa_d \, A_1 \wedge (dA_1)^{\frac{d}{2}-1} \biggr) \, \delta A_1 \,.
\ee
This can be cancelled by imposing the boundary condition
\be
\label{initial b.c. in app}
\star \, A_1 = - \frac{i}{R^2}  \underbrace{ \biggl( b_{d-1} + \frac{d}{2} \, \kappa_d \, A_1 \wedge (dA_1)^{\frac{d}{2}-1} \biggr) }_{\ds \cT_0} + \star \cA_1
\ee
and adding the boundary term
\be
S_\partial^{(0)} = - \frac{1}{4\pi R^2} \int_{\partial X_{d+1}} \biggl( b_{d-1} + \frac{d}{2} \, \kappa_d \, A_1 \wedge (dA_1)^{\frac{d}{2}-1} \biggr) \wedge \star \, \biggl( b_{d-1} + \frac{d}{2} \, \kappa_d \, A_1 \wedge (dA_1)^{\frac{d}{2}-1} \biggr) .
\ee
However, there is no gauge transformation of $\cA_1$ that makes the boundary condition gauge invariant. The only way to have a gauge-invariant boundary condition is to add terms that mix $\cA_1$ with the dynamical fields. The simplest such modification is to replace $\cT_0$ in (\ref{initial b.c. in app}) with
\be
\cT_0' = \cT_0 - \frac{d}{2} \, \kappa_d \, \cA_1 \wedge (dA_1)^{\frac{d}{2}-1} \,.
\ee
Consequently we must modify the boundary term into
\be
- \frac{1}{4\pi R^2} \int_{\partial X_{d+1}} \cT_0' \wedge \star \, \cT_0' \,.
\ee
However, since the boundary condition now imposes $\delta \cT_0' = iR^2 \star \delta A_1$, we get an extra unwanted term in the variational principle:
\be
- \frac{i}{2\pi} \int_{\partial X_{d+1}} \frac{d}{2} \, \kappa_d \, \cA_1 \wedge (dA_1)^{\frac{d}{2}-1} \wedge \delta A_1 \,.
\ee
This can be cancelled by adding a topological term proportional to $\cA_1 \wedge A_1 \wedge (dA_1)^{\frac{d}{2}-1}$ to the boundary term. Indeed 
\be
\int_{\partial X_{d+1}} \!\!\! \delta \Bigl( \cA_1 \, A_1 \, (dA_1)^{\frac{d}{2}-1} \Bigr) = \int_{\partial X_{d+1}} \Bigl( \tfrac{d}{2} \, \cA_1 \, (dA_1)^{\frac{d}{2}-1} \, \delta A_1 - \bigl( \tfrac{d}{2}-1 \bigr) \, d\cA_1 \, A_1 \, (dA_1)^{\frac{d}{2}-2} \, \delta A_1 \Bigr) \,.
\ee
However, this also produces an extra term that must be cancelled. This is easily achieved by modifying both the boundary condition and the boundary term by the addition of this extra term to $\cT_0'$. This produces 
\be
\cT_1 = \cT_0' + \kappa_d \, \bigl( \tfrac{d}{2} - 1 \bigr) \, d\cA_1 \, A_1 \, (dA_1)^{\frac{d}{2}-2} \,.
\ee
At the same time we modify the boundary term that, including the new topological term, becomes
\be
S_\partial^{(1)} = - \frac{1}{4\pi R^2} \int_{\partial X_{d+1}} \cT_1 \wedge \star \, \cT_1 + \frac{i}{2\pi} \int_{\partial X_{d+1}} \kappa_d \, \cA_1 \wedge A_1 \wedge (dA_1)^{\frac{d}{2}-1} \,.
\ee
These new boundary condition and boundary term give a consistent variational principle. However, the boundary condition is again non gauge invariant because of the last term we added to $\cT_1$, and we have to repeat the procedure above. 

At each step, the non-gauge-invariant piece in the boundary condition becomes of one lower degree in $A_1$ (and one higher in $\cA_1$). Hence, the procedure stops when we reach a term linear in $A_1$: we can make the boundary condition gauge invariant by adding a term purely in $\cA_1$, which does not modify the variational principle. The procedure stops after $(d/2-1)$ steps, yielding the boundary condition
 \be
\star \, \bigl( A_1 - \cA_1 \bigr) = - \frac{i}{R^2} \, \Bigl( \Omega_{d-1} - \kappa_d \, \cA_1 \, (\cA_1)^{\frac{d}{2}-1} \Bigr)
\ee
where 
\be
\Omega_{d-1} = b_{d-1} + \kappa_d \sum_{r=0}^{\frac{d}{2}-2} \bigl( \tfrac{d}{2} - r \bigr) \, (d\cA_1)^r \, \bigl( A_1 - \cA_1 \bigr) \, (dA_1)^{\frac{d}{2}-1-r} + \kappa_d \, (d\cA_1)^{\frac{d}{2}-1} \, A_1 \,.
\ee
The corresponding boundary term is
\be
S_\partial = - \frac{1}{4\pi R^2} \int_{\partial X_{d+1}} \Omega_{d-1} \wedge \star \, \Omega_{d-1} + \frac{i \kappa_d}{2\pi} \, \sum_{r=0}^{\frac{d}{2}-2} \, \int_{\partial X_{d+1}} \cA_1 \, (d\cA_1)^r \, A_1 \, (dA_1)^{\frac{d}{2}-r-1} \,.
\ee

As a sanity check, we can verify that the boundary theory is anomalous under $U(1)$ gauge transformations. Under $\delta A_1 = \delta \cA_1= d \lambda_0$ the topological terms on the boundary produce
\bea
\label{eq:dgtop}
& \frac{i \kappa_d}{2\pi} \, \sum_{r=0}^{\frac{d}{2}-2} \, \int_{\partial X_{d+1}} \biggl( d\lambda_0 \, (d \cA_1)^r \, A_1 \, (dA_1)^{\frac{d}{2}-r-1} + \cA_1 \, (d \cA_1)^r \, d\lambda_0 \, (dA_1)^{\frac{d}{2}-r-1} \biggr) \\
&\qquad = \frac{i \kappa_d}{2\pi} \, \sum_{r=0}^{\frac{d}{2}-2} \, \int_{\partial X_{d+1}} \lambda_0 \, \Bigl( (d\cA_1)^{r+1} (dA_1)^{\frac{d}{2}-r-1} - (d\cA_1)^r (dA_1)^{\frac{d}{2}-r} \Bigr) \\
&\qquad = \frac{i \kappa_d}{2\pi} \int_{\partial X_{d+1}} \Bigl( \lambda_0 \, (d\cA_1)^{\frac{d}{2}-1} \, (dA_1) - \lambda_0 \, (dA_1)^{\frac{d}{2}} \Bigr) \,. 
\eea
Then, using the boundary condition,
\begin{align}
\delta S_\partial &= \frac{i \kappa_d}{2\pi} \!\int_{\partial X_{d+1}} \!\!\!\!\! d\lambda_0 \, (d\cA_1)^{\frac{d}{2}-1} \bigl(A_1 - \cA_1 \bigr) - \frac{\kappa_d^2 }{2\pi R^2} \!\int_{\partial X_{d+1}} \!\!\!\!\! d\lambda_0 \, (d\cA_1)^{\frac{d}{2}-1} \wedge \star \, \bigl( (d\cA_1)^{\frac{d}{2}-1} \, \cA_1 \bigr) \\
& - \frac{ \kappa_d^2}{4\pi R^2} \!\int_{\partial X_{d+1}} \!\!\!\!\! d \lambda_0 \, (d\cA_1)^{\frac{d}{2}-1} \wedge \star \, \bigl( d \lambda_0 \, (d\cA_1)^{\frac{d}{2}-1} \bigr) + \frac{i \kappa_d}{2\pi} \!\int_{\partial X_{d+1}} \!\!\!\!\! \bigl( \lambda_0\, (d\cA_1)^{\frac{d}{2}-1} (dA_1) - \lambda_0 \, (dA_1)^{\frac{d}{2}} \bigr) \,. \nn
\end{align}    
The bulk contributes with a term
\be
\delta S = - \frac{i\kappa_d}{2\pi} \int_{\partial X_{d+1}} d\lambda_0 \, A_1 \, (dA_1)^{\frac{d}{2}-1}
\end{equation}
which, together with the last term in \eqref{eq:dgtop}, combines to a total derivative (on the boundary) and can be neglected. We remain with
\be
\delta S_\text{tot} = - \frac{i \kappa_d}{2\pi} \int_{\partial X_{d+1}} \!\!\! d\lambda_0 \, (d\cA_1)^{\frac{d}{2}-1} \, \cA_1 - \delta \biggl[ \frac{ \kappa_d^2}{4\pi R^2} \int_{\partial X_{d+1}} \Bigl( \cA_1 \, (d\cA_1)^{\frac{d}{2}-1} \Bigr) \wedge \star \, \Bigl( \cA_1 \, (d\cA_1)^{\frac{d}{2}-1} \Bigr) \biggr] \,.
\ee
We can isolate the anomalous variation adding a final counterterm
\be
S_\text{c.t.} = \frac{ \kappa_d^2}{4\pi R^2} \int_{\partial X_{d+1}} \Bigl( \cA_1 \, (d\cA_1)^{\frac{d}{2}-1} \Bigr) \wedge \star \, \Bigl( \cA_1 \, (d\cA_1)^{\frac{d}{2}-1} \Bigr) \,.
\ee

%%%%%%%%%%%%%%%%%%%%%%%%%%%%%%%%%%%%%%%%%%%%%%%%%%%%

\section{Non-compact TQFTs}
\label{app:TQFT}

In this appendix we provide a mathematical definition and details on the TQFTs with infinitely many operators introduced in \cite{Brennan:2024fgj, Antinucci:2024zjp} and used as holographic duals. The main issue is defining the theory with cutting and gluing while avoiding infinities from inserting a complete basis of states. We argue that this is possible if all manifolds have at least one non-empty boundary component. On the other hand, the partition functions on closed manifolds will be generically infinite.

\paragraph{Review of standard TQFTs.}
Recall that standard TQFTs in $d$ dimensions are defined by a symmetric monoidal functor $Z\,{:}\: \text{Bord}_d^{\text{SO}} \rightarrow \text{Vec}_\bC$ from the category of oriented bordisms to the category of complex vector spaces \cite{Atiyah:1989vu} (see \eg{} \cite{Carqueville:2017fmn} for a detailed review). A vector space \mbox{$\cH_{X_{d-1}} = Z(X_{d-1})$} is assigned to any closed codimension-one manifold and a linear map $Z(Y_d)\,{:}\: \cH_{X_{d-1}} \rightarrow \cH_{X_{d-1}'}$ to any bordism $Y_d\,{:}\: X_{d-1} \rightarrow X_{d-1}'$, namely an oriented manifold with boundary $\partial Y_d = X_{d-1} \sqcup \overline{X}_{d-1}'$ (here bar means orientation reversal) with \emph{in} and \emph{out} components given by $X_{d-1}$ and $X_{d-1}'$ respectively.%
\footnote{The same manifold, with the same orientation, can be viewed as a bordism $\overline{X}_{d-1}' \rightarrow \overline{X}_{d-1}$.}
Functoriality implies that the vector space for a disjoint union is the tensor product, and gluing $Y_d$ with $Y_d'$ along a common boundary corresponds to composing linear maps. 

In practice, it is convenient to work with an explicit basis. Hence, to concretely construct a TQFT we need the following ingredients: 
\begin{itemize}
    \item Vector spaces $\cH_{X_{d-1}}$ with a basis $|a \rangle$. We also denote by $|\bar{a} \rangle$ a basis of $\cH_{\overline{X}_{d-1}}$.
    \item For any $d$-dimensional manifold $Y_d$ with incoming and outgoing connected boundary components, respectively, $X_{d-1, \text{in}}^{i}$, $i=1,2, \ldots$ and $X_{d-1, \text{out}}^{j}$, $j=1,2, \ldots$ we assign a tensor $Z(Y_d)_{\{a_i\}, \{ b_j \}}$. This specifies the linear map $\bigotimes _{i} \cH_{\text{in},i}\rightarrow \bigotimes _j \cH_{\text{out},j}$
\be
Z(Y_d) \, \biggl(|a_1\rangle \otimes |a_2\rangle \otimes \cdots \biggr) = \sum_{b_j} Z(Y_d)_{\{ a_i \}, \{b_j \}} \, \biggl( |b_1\rangle \otimes |b_2\rangle \otimes \cdots \biggr) \,.
\ee
\end{itemize}
Notice that the vector spaces $\cH_{X_{d-1}} $ are not endowed with a scalar product as an extra datum: this simply arises from the composition of bordisms. To see this notice that, for any $X_{d-1}$, we can construct the cylinder $X_{d-1}\times [0,1]$ that can be viewed both as the \emph{straight cylinder}, namely a bordism $X_{d-1}\rightarrow X_{d-1}$, or as the \emph{horseshoe}, namely a bordism $X_{d-1}\otimes \overline{X}_{d-1}\rightarrow \emptyset$.%
\footnote{The vector space associated with the empty $(d-1)$-dimensional manifold is $\cH_{\emptyset} = \bC$.}
In the first case the functor $Z$ associates the identity map $\operatorname{Id}_{\cH_{X_{d-1}}} {:}\: \cH_{X_{d-1}} \rightarrow \cH_{X_{d-1}}$, while in the second case it gives a bilinear pairing $\eta ({X_{d-1}}) \,{:}\: \cH_{X_{d-1}}\otimes \cH_{\overline{X}_{d-1}}\rightarrow \bC$. In components these read:
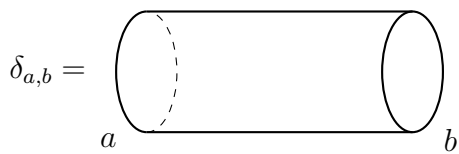
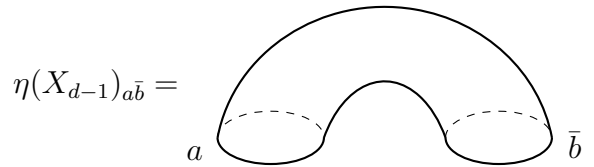
\begin{equation*}
\begin{tikzpicture}
	\node at (-3.3, 0) {\large$\delta_{a,b}=$};
	% Draw the left ellipse (top boundary)
	\node at (-2.5, -0.9) {$a$};
	\node at (2, -0.9) {$b$};
	\draw[dashed] (-2, -0.8) arc[start angle = -90, end angle = 90, x radius = 0.4, y radius = 0.8];
	\draw[thick] (-2, -0.8) arc[start angle=-90, end angle=-270, x radius = 0.4, y radius = 0.8];
	% Draw the right ellipse (bottom boundary)
	\draw[thick] (1.5, 0) ellipse[x radius = 0.4, y radius = 0.8];
	% Draw the top horizontal line
	\draw[thick] (-2, 0.8) -- (1.5, 0.8);
	% Draw the bottom horizontal line
	\draw[thick] (-2, -0.8) -- (1.5, -0.8);
\begin{scope}[shift={(8.5,0)}]
	\node at (-3.1, 0) {\large$\eta(X_{d-1}) _{a\bar{b}}=$};
	\node at (-1.8, -0.9) {$a$};
	\node at (3.2, -0.8) {$\bar{b}$};
	% Connect the right piece of the left bottom circle to the left piece of the right bottom circle
	\draw[thick] (-1.5, -0.702) arc [start angle = 170, end angle = 10, x radius = 2.234, y radius = 2.1];
	% Connect the left piece of the left bottom circle to the right piece of the right bottom circle
	\draw[thick] (-0.101, -0.704) arc [start angle = 150, end angle = 30, x radius = 0.925, y radius = 1.5];
	% Draw the bottom circles with separate solid and dashed lines
	% Left bottom circle
	\draw[thick] (-1.5, -0.7) arc[start angle = 180, end angle = 360, x radius = 0.7, y radius = 0.35];
	\draw[dashed] (-1.5, -0.7) arc[start angle = 180, end angle = 0, x radius = 0.7, y radius=0.35];
	% Right bottom circle
	\draw[thick] (1.5, -0.7) arc[start angle =180, end angle = 360, x radius = 0.7, y radius = 0.35];
	\draw[dashed] (1.5, -0.7) arc[start angle = 180, end angle = 0, x radius = 0.7, y radius = 0.35];
\end{scope}
\end{tikzpicture}
\end{equation*}
One can show that $\eta(X_{d-1})$ is a non-degenerate pairing that defines an isomorphism \mbox{$\cH_{\overline{X}_{d-1}} \!\cong \cH_{X_{d-1}}^\vee$}. This allows us to identify the basis $|\bar{a}\rangle$ of $\cH_{\overline{X}_{d-1}}\!$ with the dual basis $\langle a |$ of $\cH_{X_{d-1}}^\vee \!$ defined by $\langle a | b\rangle =\delta _{a,b}$:
\be
|\bar{b}\rangle = \sum_a \eta_{a,\bar{b}} \, \langle a | \,.
\ee

With these pieces of data, it is clear how to glue various bordisms along common boundaries to generate others. The common boundaries must have opposite orientations. When one boundary is incoming and the other one is outgoing, the gluing is just the composition. On the other hand, if both are incoming (or both outgoing), we use $\eta(X_{d-1})_{a,\bar{b}}$. More concretely, let $Y_d$ be a (possibly disconnected) bordism $\bigsqcup_i X_{d-1, \text{in}}^i \rightarrow \bigsqcup_j X_{d-1,\text{out}}^j$. If $X_{d-1,\text{in}}^1=X_{d-1,\text{out}}^1$ we can generate $\widetilde{Y}_d$ by gluing the two, and the associated tensor is
\be
Z\bigl( \widetilde{Y}_d \bigr) _{\{a_2 , \ldots\} , \{ b_2, \ldots \}} = \sum\nolimits_{a_1} \, Z(Y_d)_{ \{a_1, a_2, \ldots \}, \{a_1, b_2, \ldots \}} \,.
\ee
If instead $X_{d-1,\text{in}}^1 = \overline{X}{}_{d-1,\text{in}}^2$ the tensor associated with the manifold obtained by gluing along these boundary components is 
\be
Z\bigl( \widetilde{Y}_d \bigr)_{\{ a_3, \ldots \} , \{ b_1 \}} = \sum\nolimits_{a_1,a_2} \, Z(Y_d)_{ \{a_1, a_2, a_3, \ldots \} , \{b_1, \ldots \} } \; \eta\bigl( X_{d-1,\text{in}}^1 \bigr)_{a_1,a_2} \,.
\ee

Clearly, these pieces of data cannot be arbitrary: if the same manifold $Y_d$ can be constructed in different ways by gluing smaller pieces, the results must coincide. Once these consistency conditions are satisfied, we can compute the tensor associated with any manifold starting from those associated with the more elementary pieces. By performing enough gluings to get a closed manifold, the result is a number: the partition function. For instance, gluing the outgoing and the incoming boundary of a cylinder $X_{d-1}\times [0,1]$ we get $X_{d-1}\times S^1$, hence
\be
\label{eq:dim of H}
Z \bigl( X_{d-1} \times S^1 \bigr) = \sum\nolimits_a \delta_{a,a} = \dim \bigl( \cH_{X_{d-1}} \bigr) \,.
\ee

\paragraph{The non-compact case.}
Already the fact \eqref{eq:dim of H} suggests that in the non-compact case closed bordisms should not be included in the definition. We want to argue that, avoiding closed manifolds, there are classes of manifolds in which we can give a precise definition of the $U(1)/\bR$ BF-like theories
\be
S = \frac{i}{2\pi} \int _{\cM_d} b_{d-p-1} \wedge dA_{p} \,.
\ee

As an illustration, we consider the case of $d=2$ with $p=1$. Hence $b_0=\phi$ is a non-compact scalar, and $A$ is a $U(1)$ gauge field. The Hilbert space $\cH_{S^1}$ can be constructed by canonical quantization. We set $\cM_2=S^1\times \bR$, with $\bR$ parametrized by $t$, and split $A = \widetilde{A} + A_0^t \, dt$. Then 
\be
S = - \frac{i}{2\pi} \int_{S^1 \times \bR} \Bigl( A_0^t \, \tilde{d}\phi \wedge dt +\phi \, \partial_t \widetilde{A} \wedge dt \Bigr) \,.
\ee
We choose the temporal gauge $A_0^t=0$, and we need to impose the Gauss law $\tilde{d}\phi = 0$, namely $\phi=\phi(t)$ is independent of the spatial coordinate. Introducing 
\be
q(t) = \int_{S^1} \widetilde{A} \,, \qquad\qquad p(t) = \frac{1}{2\pi} \phi(t) \,,
\ee
we see that $q(t) \sim q(t)+2\pi$ is a periodic variable, and the action becomes
\be
S = - i \int_\bR p \, \partial_t q \, dt \,.
\ee
This is a free infinitely-massive particle on a circle of radius $2\pi$. The quantization is straightforward. We have the commutation relations
\be
[\hat{q}, \hat{p}] = i \qquad\qquad \Rightarrow \qquad\qquad e^{i \alpha \hat{p}} \cdot e^{i n \hat{q}} = e^{i\alpha n} \; e^{in \hat{q}} \cdot e^{i\alpha \hat{p}} \,.
\ee
Here $n\in \bZ$ because of the periodicity of $\hat{q}$, while $\alpha$ is a generic real number. However the operator $e^{2\pi i \hat{p}}$ commutes with the whole operator algebra, hence it is a number that we can set to $1$. Therefore the operators acting on the Hilbert space are 
\be
\widehat{\cO}_\alpha = e^{i \alpha \hat{p}} \quad\text{with}\quad \alpha \in [0,2\pi) \,, \qquad\qquad \widehat{W}_n = e^{in \hat{q}} \quad\text{with}\quad n \in \bZ \,,
\ee
with algebra
\be
\widehat{\cO}_\alpha \, \widehat{\cO}_\beta = \widehat{\cO }_{\alpha + \beta \text{ (mod $2\pi$)}} \,, \qquad
\widehat{W}_n \, \widehat{W}_m = \widehat{W}_{n+m} \,, \qquad
\widehat{\cO}_\alpha \, \widehat{W}_n = e^{i\alpha n} \; \widehat{W}_n \, \widehat{\cO}_\alpha \,.
\ee

Starting from a simultaneous eigenstate of the $\widehat{W}_n$'s such that
\be
\widehat{W}_n \, |\theta \rangle = e^{in\theta} \, |\theta\rangle \,,
\ee
using the algebra we find
\be
\widehat{\cO}_\alpha \, |\theta \rangle = |\theta -\alpha \rangle \,.
\ee
Hence we get a basis labelled by a compact continuous variable $\theta \in U(1)$. We can also use a non-compact but countable basis, starting with an eigenstate of $\widehat{\cO}_\alpha$:
\be
\widehat{\cO}_\alpha \, |k\rangle = e^{i\alpha k} \, |k\rangle  \,.
\ee
It must be $k \in \bZ$ to respect the periodicity $\alpha \sim \alpha +2\pi$. Then using the algebra we infer
\be
\widehat{W}_n \, |k \rangle = |k+n \rangle \,.
\ee
The relation between the two basis is
\be
\label{eq:change of basis}
|k\rangle = \frac1{\sqrt{2\pi}} \int_{0}^{2\pi} \! d\theta \, e^{ik\theta} \, |\theta \rangle \,, \qquad\qquad\qquad |\theta \rangle = \frac{1}{\sqrt{2\pi}} \sum _{k\in \bZ} e^{-ik\theta} \, |k\rangle \,.
\ee

Since the Hilbert space is infinite dimensional, the partition function on $T^2$ is infinite. Let us show that, on the other hand, we can consistently define a functor on the category of open oriented bordisms. In 2d the huge computational simplifications are that the only Hilbert space is $\cH_{S^1}$, and that every 2d manifold has a pair of pants decomposition. Eventually, one also needs to \emph{fill holes} by attaching a disk. Hence, on top of the horseshoe $\eta _{ab}$, the only other data one needs to assign are the disk and the pair of pants:
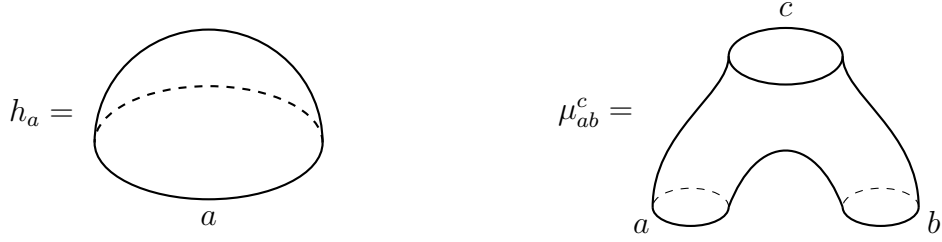
\begin{equation*}
\begin{tikzpicture}
	\node at (-2.2, 0.4) {$h_a=$};
	\node at (0,-1) {$a$};
	% Parameters for the ellipse (boundary of the hemisphere)
	\def\xradius{1.5}
	\def\yradius{0.75}
	% Draw the dashed part of the ellipse (back half)
	\draw[thick, dashed] (0:\xradius) arc[start angle=0, end angle=180, x radius=\xradius, y radius=\yradius];
	% Draw the solid part of the ellipse (front half)
	\draw[thick] (180:\xradius) arc[start angle=180, end angle=360, x radius=\xradius, y radius=\yradius];
	% Draw the hemisphere (semicircle in perspective)
	\draw[thick] (-\xradius, 0) arc[start angle=180, end angle=0, x radius=\xradius, y radius=\xradius];
\end{tikzpicture}
\qquad\qquad\qquad\qquad
\begin{tikzpicture}[scale=0.5]
	\clip (-6, -4.75) rectangle (4.2, 1.5);
	\node at (-5,-1.5) {\large$\mu_{ab}^c =$};
	% Adjust vertical distance
	\def\yshift{-4}
	\node at (-3.8, -4.5) {$a$};
	\node at (3.9, -4.4) {$b$};
	\node at (0,1.25) {$c$};
	% Draw the top circle (waist)
	\draw[thick] (0, 0) ellipse (1.5 and 0.75);
	% Connect the left bottom circle to the top circle (left side)
	\draw[thick] (-3.5, \yshift) .. controls (-3.5, -2) and (-1.5, -1) .. (-1.5, 0);
	% Connect the right bottom circle to the top circle (right side)
	\draw[thick] (3.5, \yshift) .. controls (3.5, -2) and (1.5, -1) .. (1.5, 0);
	% Connect the right piece of the left bottom circle to the left piece of the right bottom circle
	\draw[thick] (-1.5, \yshift) .. controls (-0.75, \yshift+2) and (0.75, \yshift+2) .. (1.5, \yshift);
	% Draw the bottom circles with separate solid and dashed lines
	% Left bottom circle
	\draw[thick] (-2.5, \yshift) ++(180:1) arc[start angle=180, end angle=360, x radius=1, y radius=0.5];
	\draw[dashed] (-2.5, \yshift) ++(0:1) arc[start angle=0, end angle=180, x radius=1, y radius=0.5];
	% Right bottom circle
	\draw[dashed] (2.5, \yshift) ++(0:1) arc[start angle=0, end angle=180, x radius=1, y radius=0.5];
	\draw[thick] (2.5, \yshift) ++(180:1) arc[start angle=180, end angle=360, x radius=1, y radius=0.5];
\end{tikzpicture}
\end{equation*}
The numbers $h_a$ define a distinguished state $|HH\rangle =\sum _a h_a |a\rangle$, called the Hartle--Hawking state. These two data must satisfy the obvious condition that if we fill one of the two incoming holes of the pair of pants with the Hartle--Hawking state we get the cylinder:
\be
\label{eq:HH consistency}
\sum\nolimits_b \, \mu_{ab}^c \, h_b = \delta_{a,c} \,.
\ee
The only other consistency condition is the independence from the chosen pair of pants decomposition, that reduces to the Froboenius condition \cite{Abrams:1996ty}:
\be
\label{eq:froboenius}
\sum\nolimits_c \, \mu_{a,b}^c \, \mu_{c,d}^e = \sum\nolimits_c \, \mu_{a,c}^e \, \mu_{b,d}^c \,.
\ee

Let us use the continuous basis $|\theta \rangle$. The cylinder (identity) becomes a delta function $\delta(\theta_1 -\theta _2)$. Moreover, we define
\be
\label{eq:data continuous}
h_\theta = \delta(\theta) \,, \qquad\qquad \eta_{\theta _1,\theta _2} = \delta (\theta_1 + \theta_2) \,, \qquad\qquad \mu_{\theta _1, \theta_2}^{\theta_3} = \delta (\theta_1 + \theta_2 - \theta_3) \,.
\ee
Also, all sums are replaced by integrals on $[0,2\pi)$ in this basis. The condition \eqref{eq:HH consistency} is obviously satisfied, while the Froboenius condition (\ref{eq:froboenius}) reads 
\be
\int_0^{2\pi} \! d\theta \; \delta (\theta_1 + \theta_2 - \theta) \, \delta (\theta + \theta_3 - \theta_4) = \int_0^{2\pi} \! d\theta \; \delta(\theta_1 + \theta - \theta_4) \, \delta (\theta_2 + \theta_3 - \theta)
\ee
which is satisfied since both sides are equal to $\delta (\theta_1 + \theta_2 + \theta_3 - \theta_4)$. The choice of these data is motivated by the fact that the continuous basis $|\theta\rangle$ is related, by the state/operator correspondence, with the local operators $\cO_\alpha (x)=e^{i\frac{\alpha}{2\pi}\phi(x)}$, and the pair of pants must reproduce their OPE $\cO_\alpha \, \cO _\beta = \cO_{\alpha +\beta}$. Then the Hartle--Hawking state is fixed by \eqref{eq:HH consistency}.

With these pieces of data, we can compute the value of the functor for arbitrary bordisms with a non-empty boundary. The simplest nontrivial such manifold is the torus with a puncture. This can be obtained from the pair of pants by gluing one of the two incoming boundaries with the outgoing one. Denoting by $\theta$ the label of the puncture, namely the non-glued circle, the result is%
\footnote{We denote a genus $g$ Riemann surface as $\Sigma_g$.}
\be
Z \bigl( \Sigma_1 \smallsetminus P_\theta \bigr) = \int_{0}^{2\pi} \! d\theta' \; \delta(\theta) = 2 \pi \, \delta (\theta) \,.
\ee
This is a projector on the Hartle--Hawking state. Another simple example is the torus with two punctures that can be obtained from the previous result by gluing the remaining boundary to the outgoing boundary of another pair of pants. Hence, the result is 
\be
Z \Bigl( \Sigma_1 \smallsetminus \bigl\{ P_{\theta_1}, P_{\theta_2} \bigr\} \Bigr) = \int_0^{2\pi} \! d\theta' \; \delta (\theta_1 + \theta_2 - \theta') \, 2\pi \delta(\theta') = 2\pi \, \delta (\theta_1 + \theta_2)  \,.
\ee
We can now put these two examples together, gluing the boundary of a torus with one puncture to one of the two boundaries of the torus with two punctures, resulting in a genus-two surface with a puncture:
\be
Z \bigl( \Sigma_2 \smallsetminus P_\theta \bigr) = \int_0^{2\pi} \! d\theta' \; 2\pi \delta (\theta + \theta') \, 2\pi \delta(\theta') = (2\pi)^2 \, \delta(\theta) \,.
\ee
Proceeding in this way it is not hard to prove the general result. The value of the functor an a genus $g$ surface with $n$ incoming boundaries labelled by $\theta_1, \ldots, \theta_n$ and $m$ outgoing boundaries labelled by $\theta_1', \ldots , \theta_m'$ is given by
\be
Z \Bigl( \Sigma_g \smallsetminus \bigl\{ P_{\theta_1}, \ldots , P_{\theta_n}, P_{\theta_1'}, \ldots , P_{\theta_m'} \bigr\} \Bigr) = (2\pi )^g \; \delta \bigl( \theta_1 + \ldots + \theta_n - \theta_1' - \ldots - \theta_m' \bigr) \,.
\ee

The important observation is that the partition function on compact Riemann surfaces is infinite. Indeed, a compact Riemann surface of genus $g$ is obtained by closing the hole of a one-punctured Riemann surface $\Sigma_g \smallsetminus P_\theta$ by means of gluing the Hartle--Hawking state. The result is clearly infinite:
\be
Z(\Sigma_g) = \int_0^{2\pi} \! d\theta \; (2\pi)^g \, \delta(\theta) \, \delta(\theta) = (2\pi)^g \, \delta(0) \,.
\ee
We conclude that the TQFT is well defined on the category of open oriented bordisms.

Let us remark that, given the Hilbert space we constructed, there is another set of data that can be formulated, which is essentially the same as the one we discussed but in the discrete basis $|k\rangle$:
\be
\label{eq:data discrete}
h'_k = \delta_{k,0} \,, \qquad\qquad \eta'_{k_1, k_2} = \delta_{k_1, -k_2} \,, \qquad\qquad (\mu')_{k_1, k_2}^{k_3} = \delta_{k_1+k_2, k_3} \,.
\ee
With these data one gets infinite answers even on open manifolds, as soon as they have a non-trivial topology. It must be noticed that, indeed, these are not merely the data \eqref{eq:data continuous} written in a different basis: translating \eqref{eq:data continuous} in the discrete basis using (\ref{eq:change of basis}) we get
\be
h_k = \frac{1}{\sqrt{2\pi}} \,, \qquad\qquad \eta_{k_1, k_2} = \delta_{k_1, k_2} \,, \qquad\qquad \mu_{k_1, k_2}^{k_3} = \sqrt{2\pi} \; \delta_{k_1,k_2} \, \delta_{k_1, k_3} \,.
\ee
We conclude that \eqref{eq:data continuous} and \eqref{eq:data discrete} really define two different TQFTs.

How did we choose one instead of the other? As we already pointed out, in 2d TQFT the choice is really dictated by the fact that the pair of pants is related with the OPE of local operators. The data \eqref{eq:data discrete} would then be relevant for the TQFT with Lagrangian formulation
\be
S' = \frac{i}{2\pi} \int_{\cM_2} \Phi \, da_1 \,,
\ee
where $\Phi \sim \Phi + 2\pi $ is a compact scalar, while $a_1$ an $\bR$ gauge field. Canonical quantization produces the same Hilbert space as the theory with non-compact scalar and $U(1)$ gauge field; however, here the local operators $\cO_n(x)=e^{in\Phi(x)}$ are labeled by an integer, and hence are related with the discrete basis by the state/operator correspondence. For this reason, in contrast to the previous case, the quantization of this theory produces the data \eqref{eq:data discrete} in which the pair of pants gives the Abelian fusion algebra in the discrete basis.

\bibliographystyle{ytphys}
\baselineskip=0.90\baselineskip
\bibliography{TopGravity}

\end{document}